\documentclass{aa}  

\usepackage{natbib}

\usepackage{graphicx}
\usepackage[varg]{txfonts}
\usepackage{lipsum}
\usepackage{subcaption}         
\usepackage{lscape}             
\usepackage{placeins}           
                                
\usepackage{physics}
\usepackage{xcolor}
\usepackage{siunitx}
\newcommand{\msol}{\mathrm{M}_\odot}
\newcommand{\tensorfont}[1]{\boldsymbol{\mathsf{#1}}}

\RequirePackage{xcolor}

\providecommand{\DIFdel}[1]{}

\providecommand{\DIFdelFL}[1]{}

\begin{document}

   \title{CRESCENDO II: Spectral cosmic rays with improved energy losses and realistic supernova seeding}

%

   \author{Daniel Karner \inst{1}\corrauth{dkarner@usm.lmu.de}
        \and Ludwig M. B{\"o}ss\inst{2}\email{lboess@uchicago.edu}
        \and Klaus Dolag\inst{1,3}\email{kdolag@MPA-Garching.MPG.DE}
        \and Ildar Khabibullin\inst{4,1,3}\email{ildar@usm.lmu.de}
        }

   \institute{
   Universit{\"a}ts-Sternwarte, Fakult{\"a}t f{\"u}r Physik, Ludwig-Maximilians-Universit{\"a}t M{\"u}nchen, Scheinerstr. 1, 81679 M{\"u}nchen, Germany
   \and Department of Astronomy and Astrophysics, The University of Chicago, William Eckhart Research Center, 5640 S. Ellis Ave. Chicago, IL 60637, USA
   \and Max-Planck-Institut f{\"u}r Astrophysik, Karl-Schwarzschild-Straße 1, 85741 Garching, Germany
    \and
         Rudolf Peierls Centre for Theoretical Physics, Department of Physics, University of Oxford, Clarendon Laboratory, Parks Rd, Oxford, OX1 3PU, United Kingdom
         }
   \date{Received \today}

 
  \abstract
   {Cosmological simulation codes with subgrid models for cosmic rays (CRs) help us better understand their impact on baryonic feedback and non-thermal radiation in galaxies and galaxy clusters. An accurate numerical description requires a spectrally resolved treatment of the CR population, because virtually all transport, acceleration and loss processes depend on energy.}
   {We advance the treatment of CR electrons and protons in the on-the-fly spectral CR solver CRESCENDO in \textsc{OpenGadget3}.}
   {We implement several new energy loss processes for both protons and electrons and improve the computation of their energies and pressures beyond the ultra-relativistic approximation. Moreover, we present a subgrid model for CR seeding by supernova remnants, in which physically motivated spectra are injected at sites of ongoing star formation.}
   {We test the newly implemented loss processes and the coupling between CR injection and star formation in idealized setups. We also highlight numerical subtleties, such as the differences arising when hadronic losses are modelled as continuous or catastrophic process, and the advantages of using a flexible spectral cut-off and abandoning the ultra-relativistic approximation. Furthermore, we show that using analytical approximations to compute energy fluxes can cause the slope reconstruction to fail.}
   {Future applications of our spectral cosmic-ray model in large-scale, full-physics cosmological simulations will represent an important step towards building a robust and observationally verifiable link between the microphysical and macrophysical aspects of the CR component in the modern paradigm of galaxy evolution.}


   \maketitle
    \nolinenumbers

\section{Introduction}

Cosmic rays (CRs) are relativistic electrons, protons and heavier nuclei that permeate the interstellar medium (ISM) and the intergalactic medium (IGM). In typical spiral galaxies their energy density is comparable to that of magnetic fields, electromagnetic radiation and thermal gas \citep{Beck2015}, making them an important ingredient of baryonic feedback \citep{2013ApJ...777L..38H}. Consequently, there has been a growing interest over the past decade in incorporating them into cosmological simulation codes. Even the simplest subgrid models for CRs differ qualitatively from standard feedback prescriptions for supernovae or active galactic nuclei, where a certain amount of thermal energy is injected locally and/or momentum is imparted to the gas \citep{Hopkins.etal2023}. This difference arises, because CR protons have long cooling times compared with the hot gas involved in thermal feedback, their complicated spatial propagation is a combination of advection, streaming and anisotropic diffusion and they effectively couple to the background plasma. Together, these properties allow CRs to influence galaxy evolution, especially with respect to launching galactic winds, (mildly) suppressing the star formation rate and shaping the structure of the circumgalactic medium \citep[for recent reviews see][]{Ruszkowski.Pfrommer2023, Owen.etal2023, Hopkins2026}.

In addition, both CR protons and CR electrons are responsible for several non-thermal radiation processes on various scales. In star-forming galaxies CRs accelerated by supernova remnants give rise to the well-known correlations between the far-infrared emission, that is proportional to the star formation rate, and the $\gamma$-ray emission, caused by decay products of hadronic interactions; and the far-infrared and radio emission, generated by relativistic electrons \citep[see][and references therein]{Ruszkowski.Pfrommer2023}. In galaxy clusters, where CR feedback is less important, synchrotron-emitting CR electrons can fill large volumes, giving rise to structures like radio halos and radio relics. These phenomena provide important diagnostics for CR acceleration and transport in the intracluster medium \citep[for reviews see][]{Brunetti.Jones2014, Bykov.etal2019, vanWeeren.etal2019, Vazza.Botteon2024}.

To improve our understanding of cosmic-ray feedback on galactic and cluster scales and to model the associated non-thermal emission for direct comparison with observations, different subgrid models for cosmic rays are now available in various simulation codes. There are two main approaches. First, ``one-bin'' or ``grey'' models treat CRs as a relativistic fluid, where moments of the distribution function, like the integrated CR energy density, are evolved. This description of CRs has been included in Gadget \citep{2006MNRAS.367..113P}, Piernik \citep{2013ApJ...777L..38H}, Athena \citep{Jiang.Oh2018}, Athena++ \citep{Zhao.etal2026}, Gizmo \citep{Chan.etal2019}, Arepo \citep{Pfrommer.etal2017, Thomas.Pfrommer2019}, Enzo \citep{Vazza.etal2012} and Ramses \citep{Rosdahl.etal2025}. Second, in spectral models the distribution functions of all cosmic ray species are evolved either on-the-fly or in post-processing, which has been implemented in Gadget
\citep{2013MNRAS.429.3564D, Boess.etal2023}, Arepo \citep{Winner.etal2019, Girichidis.etal2020, Girichidis.etal2022}, Enzo \citep{Vazza.etal2016, Wittor.etal2017}, Gizmo \citep{Hopkins.etal2022a}, Piernik \citep{Ogrodnik.etal2021, BaldacchinoJordan.etal2025} and Ramses \citep{Diallo.etal2026}. 

Although these codes employ somewhat different prescriptions for CR transport and energy losses, tailored to their respective applications, they share the common goal of capturing the dynamical and radiative impact of CRs. In this paper we present updates to the spectral CR model CRESCENDO \citep{Boess.etal2023} implemented in the cosmological simulation code \textsc{OpenGadget3} \citep{2026arXiv260917658D}. Our primary aim is to model non-thermal emission in cosmological simulations. To this end, we incorporate additional energy-loss processes, improve the calculation of CR energy and pressure, and introduce template spectra for CR injection by supernova remnants. The underlying numerical methods are explained in detail in Sec.~\ref{sec: Numerical methods} and tested in idealized setups in Sec.~\ref{Sec: Numerical tests}.

\section{Numerical methods}\label{sec: Numerical methods}

\begin{figure*}
    \includegraphics[width=\textwidth]{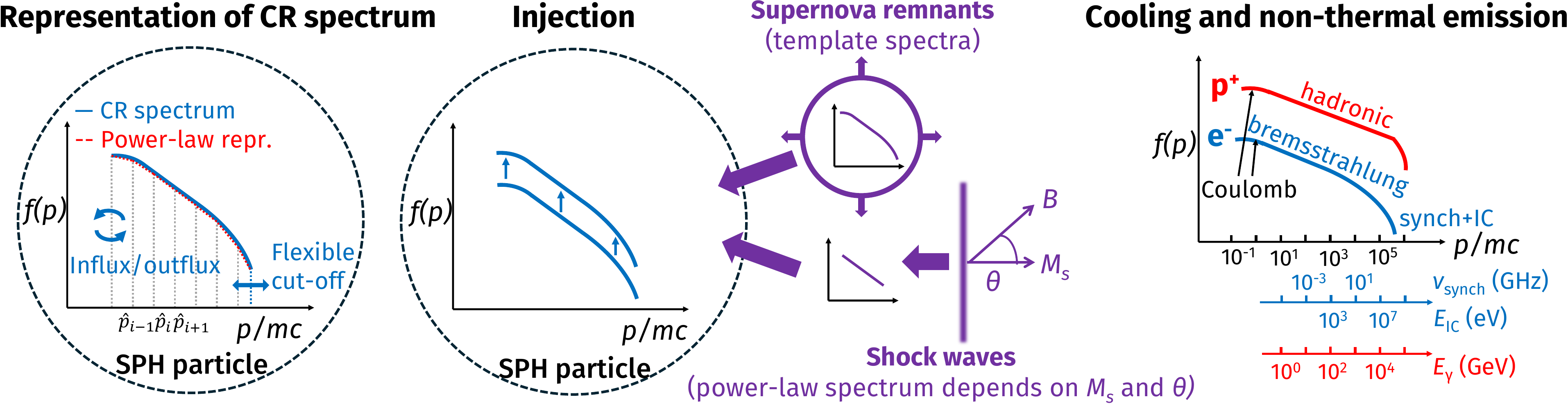}
    \caption{Summary of the most important points discussed in the main text, namely 1) the numerical representation of the CR spectra as piecewise power-laws with influx/outflux through the lowest bin and a flexible upper cut-off, 2) the injection of CRs via supernova remnants (model spectra) and shock waves (slope and normalization depend on Mach number and orientation of magnetic field), 3) a summary of relevant energy loss processes, in which momentum range they dominate and the energy bands of the corresponding non-thermal emission (electrons: radio and X-ray emission from synchrotron and IC losses; protons: $\gamma$-rays from hadronic interactions).}
    \label{fig: overview of numerical method}
\end{figure*}

\subsection{Discretization of the distribution function}
In \textsc{OpenGadget3} we couple all CR physics to smoothed particle hydrodynamics (SPH). We attach a spectrum for CR protons and for CR electrons to every SPH particle and evolve it in time using the on-the-fly Fokker-Planck solver CRESCENDO \citep{Boess.etal2023} that is based on the two-moment approach by \citet{Miniati2001}. Instead of using the full $6+1$-dimensional CR distribution function $F(\vb{x}, \vb{p}, t)$, we only consider its isotropic part $f(\vb{x}, p, t)$, which only depends on the absolute value of the particle momentum $p$. The momentum range $[p_\mathrm{min}, p_\mathrm{max}]$ for which $f$ is non-zero is set at the beginning of a simulation and divided into $N_\mathrm{bins}$ bins, which are equally spaced on a $\lg(p)$-axis. It is natural to represent the distribution function as a piecewise power-law in momentum space, because real cosmic-ray spectra indeed resemble power-laws over a large dynamical range. Therefore, in each bin (denoted by an index $i$) $f$ is characterized by a slope $q_i$ and a spectral normalization $f_i(\vb{x},t)$, whose spatial dependence is determined by the location of the SPH particle: 
\begin{equation}
\begin{aligned}
    f(\vb{x}, p, t)\big|_{p \in [p_i,p_{i+1}]} &= f_i(\vb{x}, t) \left(\frac{p}{p_i} \right)^{-q_i},  \; i \in \{1, \ldots, N_\mathrm{bins}\}\,, \\
    \mathrm{supp}(f) &= \{p: p_\mathrm{min} \leq p \leq p_\mathrm{cut}(t)\}\,.
\end{aligned}
    \label{eq: piecewise power-law representation of f}
\end{equation}
Note that we do not enforce the continuity of $f(\vb{x}, p, t)$. We fix the lower boundary $p_\mathrm{min}$ and allow an influx and outflux of CRs, whereas the spectral cut-off $p_\mathrm{cut}(t)$ may vary over time. It is initialized as $p_\mathrm{cut}(t_\mathrm{start}) = p_\mathrm{max}$ at the beginning of a simulation and recomputed after each timestep. This flexible bin boundary not only makes the representation of the distribution function more realistic, but also prevents unphysical spectral curvature, as discussed in App.~\ref{app: Importance of updating the spectral cut-off}.

Instead of evolving the normalization $f_i$ and slope $q_i$ directly, we update the CR number $n_{\mathrm{CR},i}$ and kinetic energy $\varepsilon_{\mathrm{CR},i}$ per unit mass\footnote{For our Lagrangian code, a mass discretization is more natural than evolving densities of the CR number and energy.} in every bin and then reconstruct $f_i$ and $q_i$. This so-called two-moment approach has several advantages, like numerical stability and strict conservation of the CR number and energy, and is used in various studies such as \cite{Miniati2001, Girichidis.etal2020, Ogrodnik.etal2021, Hopkins.etal2022a}. For the $i$-th momentum bin, $n_{\mathrm{CR},i}$ and $\varepsilon_{\mathrm{CR},i}$ are given by the following integrals:
\begin{align}
    n_{\mathrm{CR},i}(\vb{x},t) &= \frac{4 \pi}{\rho} \int \limits_{p_i}^{p_{i+1}} f(\vb{x}, p, t) p^2 \dd{p}\,, \label{eq: CR number integral} \\
	\varepsilon_{\mathrm{CR},i}(\vb{x},t) &= \frac{4 \pi}{\rho} \int \limits_{p_i}^{p_{i+1}} f(\vb{x}, p, t) T(p) p^2 \dd{p}\,. \label{eq: CR energy integral}
\end{align}

\subsection{Moments of the distribution function}

In previous works \citep{Boess.etal2023, Boess.etal2024, Boess.etal2025, Diesing.etal2026} the kinetic/total particle energy $T(p)\; /\; E(p)$ and the total CR pressure $P_\mathrm{CR}$ were computed in the ultra-relativistic limit $p/mc \gg 1$, i.e. $E(p) \approx T(p) \approx pc$ and $P_\mathrm{CR} = (\gamma_\mathrm{CR}-1) \varepsilon_\mathrm{CR} \rho$, where $\varepsilon_\mathrm{CR} = \sum_{i=1}^{N_\mathrm{bins}} \varepsilon_{\mathrm{CR},i}$ is the total CR energy (per unit mass) and $\gamma_\mathrm{CR}=4/3$ is a fixed adiabatic index. We improve upon this in the presented work, by using the expressions
\begin{equation}
    E(p) = mc^2 \sqrt{(p/mc)^2+1}\,, \quad T(p) = E(p) - mc^2\,,
    \label{eq: relativistic (kinetic) energy}
\end{equation}
which are also valid in the mildly relativistic regime $p/mc \lesssim 1$. This improvement is especially important for momentum loss processes that are most efficient for small $p$, like Coulomb losses (see Sec.~\ref{sec: Energy loss processes}), where the ultra-relativistic approximation becomes inaccurate. It also changes how CR spectra with a given energy content are normalized, which is relevant for the computation of non-thermal synchrotron emission from the electron spectra.

On the other hand, the total CR pressure $P_\mathrm{CR}= \sum_{i=1}^{N_\mathrm{bins}} P_{\mathrm{CR},i}$ is the sum of the pressures from all bins, given by:
\begin{equation}
    \begin{aligned}
        P_{\mathrm{CR},i}(\vb{x}, t) &= \frac{4 \pi}{3} \int \limits_{p_i}^{p_{i+1}} f(\vb{x},p,t) \,p^4 c^2\, \frac{\dd{p}}{E(p)}\,.
    \end{aligned}
    \label{eq: CR pressure integral}
\end{equation}
The energy and pressure integrals can be slightly simplified by inserting the piecewise power-law representation of the distribution function \eqref{eq: piecewise power-law representation of f} into Eqs.~\eqref{eq: CR energy integral} and \eqref{eq: CR pressure integral} and using the dimensionless momentum $\hat{p} \equiv p/mc$, where $m$ is the mass of the particle species (so either $m_\mathrm{e}$ or $m_\mathrm{p}$):
\begin{align}
    n_{\mathrm{CR},i} &= \frac{4 \pi (mc)^3 f_i \hat{p}_i^{q_i}}{\rho} \int \limits_{\hat{p}_i}^{\hat{p}_{i+1}} \hat{p}^{2-q_i} \dd{\hat{p}}\,, 
    \label{eq: CR number integral final}\\
	\varepsilon_{\mathrm{CR},i} &= \frac{4 \pi c (mc)^4 \hat{p}_i^{q_i} f_i}{\rho} \int \limits_{\hat{p}_i}^{\hat{p}_{i+1}} \left(\sqrt{\hat{p}^2 + 1} - 1 \right) \hat{p}^{2-q_i} \dd{\hat{p}} \,,
    \label{eq: CR energy integral final}\\
    P_{\mathrm{CR},i} &= \frac{4 \pi}{3} f_i c (mc)^4 \hat{p}_i^{q_i} \int \limits_{\hat{p}_i}^{\hat{p}_{i+1}} \frac{\hat{p}^{4-q_i}}{\sqrt{\hat{p}^2 + 1}} \dd{\hat{p}}\,. 
    \label{eq: CR pressure integral final}
\end{align}
Here (and in the following) we suppress the explicit dependence on $\vb{x}$ and $t$ to simplify the notation. After updating $\varepsilon_{\mathrm{CR},i}$ and $n_{\mathrm{CR},i}$, as outlined in the following sections, we update the spectral normalization $f_i$ by inverting Eq.~\eqref{eq: CR number integral final}:
 \begin{align*}
     f_i = \frac{n_{\mathrm{CR},i}\, \rho}{4 \pi (mc)^3 \hat{p}_i^{q_i}} \left( \int \limits_{\hat{p}_i}^{\hat{p}_{i+1}} \hat{p}^{2-q_i} \dd{\hat{p}} \right)^{-1}\,.
 \end{align*}
The spectral slope is updated by taking the ratio
\begin{align}
	\frac{\varepsilon_{\mathrm{CR},i}}{n_{\mathrm{CR},i}} = mc^2\, \int \limits_{\hat{p}_i}^{\hat{p}_{i+1}} \quantity(\sqrt{\hat{p}^2 + 1} - 1) \hat{p}^{2-q_i} \dd{\hat{p}} \Bigg/ \int \limits_{\hat{p}_i}^{\hat{p}_{i+1}} \hat{p}^{2-q_i} \dd{\hat{p}} 
    \label{eq: update of spectral slope}
\end{align}
and solving it for the slope $q_i$ with Brent's method. In order to numerically compute all integrals listed so far we approximate the integrands as a power-law (if they do not already have such a form). More details can be found in App.~\ref{app: Numerical computation of improved energy and pressure integrals}.

An important application of Eq.~\eqref{eq: CR pressure integral final} is the computation of the adiabatic sound speed $c_s = \sqrt{(\partial P_\mathrm{tot}/ \partial \rho)_s}$, which enters the signal velocity and the Courant-like timestep criterion. The total pressure $P_\mathrm{tot}$ is the sum of the pressures of the thermal gas, CR protons and CR electrons, i.e. $P_\mathrm{tot} = P_\mathrm{th} + P_\mathrm{CR,e} + P_\mathrm{CR,p}$, which are given by the adiabatic equation of state: 
\[ P_\mathrm{th} = K_\mathrm{th} \rho^{\gamma_\mathrm{th}}\,, \quad P_{\mathrm{CR}} = \sum_{i=1}^{N_\mathrm{bins}} P_{\mathrm{CR},i} = \sum_{i=1}^{N_\mathrm{bins}} K_{\mathrm{CR},i} \rho^{\gamma_{\mathrm{CR},i}} \,. \]
The total CR pressure for both protons and electrons is the sum of the pressure contributions from all momentum bins, each having a different adiabatic index $\gamma_{\mathrm{CR},i}$. Hence, the adiabatic sound speed is \citep[cf.][]{Girichidis.etal2022}
\begin{align}
    c_s^2 = \left( \pdv{P_\mathrm{tot}}{\rho} \right)_s = \frac{1}{\rho} \left(\gamma_\mathrm{th} P_\mathrm{th} + \sum_{i=1}^{N_\mathrm{bins}} \gamma_{\mathrm{CR},i} P_\mathrm{CR,i } \right)\,,
\end{align}
where the derivative is for a constant specific entropy $s$ and the prefactors $K$ were eliminated. For a monatomic ideal gas $\gamma_\mathrm{th} =5/3$, whereas the adiabatic indices $\gamma_{\mathrm{CR},i}$ are obtained from the integrals~\eqref{eq: CR energy integral final} and \eqref{eq: CR pressure integral final} by inverting the general ideal gas equation $P = (\gamma-1) \varepsilon \rho$:
\begin{align*}
    \gamma_{\mathrm{CR},i} &= 1 + \frac{P_{\mathrm{CR},i}}{\rho\, \varepsilon_{\mathrm{CR},i}}\\
    &= 1 + \frac{1}{3} \int \limits_{\hat{p}_i}^{\hat{p}_{i+1}} \frac{\hat{p}^{4-q_i}}{\sqrt{\hat{p}^2 + 1}} \dd{\hat{p}} \Bigg/ \int \limits_{\hat{p}_i}^{\hat{p}_{i+1}} \left(\sqrt{\hat{p}^2 + 1}-1 \right) \hat{p}^{2-q_i} \dd{\hat{p}} \,. 
\end{align*}
As expected, $\gamma_{\mathrm{CR},i} \rightarrow 4/3$ in the ultra-relativistic limit $\hat{p} \gg 1$.

\subsection{Two-moment-approach for evolving the CR spectrum}

In our previous work \citep{Boess.etal2023} we used the evolution equation~\eqref{eq: CR transport equation} for the CR distribution function to derive how $n_{\mathrm{CR},i}$ and $\varepsilon_{\mathrm{CR},i}$ change over time. For convenience, the main details are summarized in App.~\ref{app: Summary of two-moment approach}. In the following sections we will ignore the transport and source terms and only focus on the loss processes. We will also drop the subscript ``CR'' for brevity.

\subsubsection{Inclusion of loss processes}
 
The evolution of the CR number per mass only due to momentum loss processes follows directly from Eq.~\eqref{eq: dn/dt full}:
\begin{align}
	\dv{n_i}{t} = \frac{1}{\rho}\left[ -\dv{p}{t} 4 \pi p^2 f \right]^{p_{i+1}}_{p_i} \,.
    \label{eq: dn/dt}
\end{align}
It can be solved by integrating both sides from $t_0$ to $t_0 + \Delta t$, where $t_0$ is an arbitrary initial time and $\Delta t$ is the simulation timestep (see discussion below):
\begin{align}
	n_i(t_0 +\Delta t) - n_i(t_0) &= \frac{1}{\Bar{\rho}} \left(\Phi_{n,i+1} - \Phi_{n,i} \right)\,. \label{eq: n(t+dt)-n(t)}
\end{align}
Here, $\bar{\rho}$ is the mean gas density averaged over $\Delta t$,
\begin{align*}
    \bar{\rho} = \frac{1}{\Delta t} \int \limits_{t_0}^{t_0 + \Delta t} \rho(t) \dd{t} \approx \frac{\rho(t_0 + \Delta t) + \rho(t_0)}{2} \equiv \frac{\rho +\rho_\mathrm{old}}{2} \,,
\end{align*}
where $\rho_\mathrm{old}$ is the initial density. Moreover, $\Phi_{n,i}$ is defined as
\begin{equation}
	\Phi_{n,i} = \hspace{-0.2cm} \int \limits_{t_0}^{t_0 + \Delta t} \left(- \dv{p}{\tau}\right)4 \pi p^2 f(p, \tau) \big|_{p_i} \dd{\tau} 
	= \hspace{-0.3cm} \int \limits_{p_i = p(t_0 + \Delta t)}^{p_u = p(t_0)} \hspace{-0.2cm} 4 \pi p^2 f(p,t_0) \dd{p}
    \label{eq: CR number flux}
\end{equation}
and often referred to as ``CR number flux''. In fact, $\Phi_{n,i}/\bar{\rho}$ is the CR number per unit mass in the interval $[p_i, p_u]$ that flows from the current bin to the lower bin if we consider a momentum loss process with $p_u>p_i$. Hence, Eq.~\eqref{eq: n(t+dt)-n(t)} simply states that the net change in particle number within the $i$-th bin is the difference between the influx at one boundary and the outflux at the other one. In contrast, for a momentum gain (e.g. an adiabatic compression) $p_u < p_i$ and the integration interval $[p_u, p_i]$ lies in the lower bin with index $i-1$. Therefore, one has to make a case distinction for the distribution function to be used in the flux integral~\eqref{eq: CR number flux}: 
\begin{align}
    f(p, t_0) = \begin{cases}
        f_i \left(\frac{p}{p_i} \right)^{-q_i} &\quad \mathrm{for}\; p_u > p_i \quad (\mathrm{bin}\;i)\,,\\
        f_{i-1} \left(\frac{p}{p_{i-1}} \right)^{-q_{i-1}} &\quad \mathrm{for}\; p_u < p_i  \quad (\mathrm{bin}\;i-1)\,,
    \end{cases}
    \label{eq: case distinction for distribution function}
\end{align}
Hence, the flux integral~\eqref{eq: CR number flux} becomes
\begin{equation*}
\begin{aligned}
	\Phi_{n,i} 
    &=  \begin{cases} 4 \pi (mc)^3 f_i \hat{p}_{i}^{q_i} \int \limits_{\hat{p}_i}^{\hat{p}_u}  \hat{p}^{2- q_i} \dd{\hat{p}} \quad & \; \hat{p}_u > \hat{p}_i \,,\\ 
    4 \pi (mc)^3 f_{i-1} \hat{p}_{i-1}^{q_{i-1}} \int \limits_{\hat{p}_i}^{\hat{p}_u} \hat{p}^{2- q_{i-1}} \dd{\hat{p}} \quad & \; \hat{p}_u < \hat{p}_i \,.\end{cases}
\end{aligned}
\end{equation*}

The initial momentum boundary $p_u$ can be found by solving the ordinary differential equation (ODE)
\begin{equation}
    \dv{p}{t} = -b(p) \,, \quad p(t_0 + \Delta t)=p_i\,, \quad p(t_0) = p_u\,.
    \label{eq: IVP for b(p)}
\end{equation}
This is a formal initial value problem, because $p_u$ is unknown. 
However, with the variable transformation $s(t) = \Delta t -t$ this can be rewritten as
\begin{align*}
\dv{p}{s} = - \dv{p}{t} = b(p)\,, \quad p(s=0) = p_i\,, \quad p(s= \Delta t) = p_u\,,
\end{align*}
where the initial momentum $p_i$ is known and $t_0 = 0$ without loss of generality. Since $b(p)$ does not explicitly depend on time, one can use separation of variables to find an implicit solution:
\begin{align}
	\Delta t 
    = -  \int \limits_{t_0}^{t_0 + \Delta t} \frac{\dot{p}(t)}{b(p(t))} \dd{t} = - \int \limits_{p(t_0)}^{p(t_0 + \Delta t)} \frac{\dd{p} }{b(p)}= - \int \limits_{p_u}^{p_i} \frac{\dd{p}}{b(p)}\,.
    \label{eq: implicit definition of pu}
\end{align}
One can either solve this numerically using a root finder or compute it directly if the initial momentum can be explicitly written as a function $p_u = p_u(p_i, \Delta t)$.

So far, we considered only a single loss mechanism, but in general the total momentum loss rate $-\dot{p}_\mathrm{tot} = b_\mathrm{tot} = \sum_j b_j$ is the sum of several loss processes $b_j$. From the definition of the CR number flux~\eqref{eq: CR number flux} it follows immediately that the total flux is just the sum of all individual fluxes, i.e. $\Phi_{n,i} = \sum_j \Phi_{n,i,j}$ with
\begin{align*}
	\Phi_{n,i,j} &= \int \limits_{p_i}^{p_{u,j}} 4 \pi p^2 f(p,t) \dd{p}\,.
\end{align*}
For the $i$-th momentum bin the lower boundary $p_i$ is of course always fixed, whereas the upper boundary $p_{u,j}$ is computed separately for every $b_j$ using Eq.~\eqref{eq: implicit definition of pu}.

The evolution equation for $\varepsilon_{\mathrm{CR},i}$ follows from Eq.~\eqref{eq: de/dt full}:
\begin{align}
	\dv{\varepsilon_i}{t} = \left[ -\left( \dv{p}{t} \right)_\mathrm{tot} \frac{4 \pi}{\rho} p^2 T(p) f \right]^{p_{i+1}}_{p_i} - R_i \varepsilon_i \,.
    \label{eq: de/dt}
\end{align}
It appears similar to Eq.~\eqref{eq: dn/dt} apart from the additional term $R_i \varepsilon_i$ that can be interpreted as relative energy change (see App.~\ref{app: Summary of two-moment approach}):
\begin{equation}
	R_i(\vb{x},t) = \frac{1}{\varepsilon_i \, \rho} \int \limits_{p_i}^{p_{i+1}} b_\mathrm{tot}(p) \, 4 \pi p^2 f(p,t) \dv{T(p)}{p} \dd{p} \,. 
    \label{eq: relative energy change term R}
\end{equation}
Integration over the time interval $\Delta t$ leads to
\begin{equation}
    \begin{aligned}
	\varepsilon_{i}(t_0+\Delta t) - \varepsilon_{i}(t_0) &=  \frac{1}{\bar{\rho}}\left( \Phi_{\varepsilon,i+1} - \Phi_{\varepsilon, i} \right) - \int \limits_{t_0}^{t_0 + \Delta t} R_i(t) \varepsilon_{i}(t) \dd{t}\\
    &\approx \frac{1}{\bar{\rho}} \left( \Phi_{\varepsilon,i+1} - \Phi_{\varepsilon, i} \right) - \Delta t R_i(t_0) \varepsilon_i(t_0)\,.
\end{aligned}
\label{eq: updated e}
\end{equation}

The simple approximation for the integral over the relative energy change $R_i$ can be justified, because its only time-dependence emerges from the environmental parameters in $b_\mathrm{tot}$ that are constant during a simulation timestep $\Delta t$. Analogously to $\Phi_{n,i}$ we define ``CR energy fluxes'' as
\begin{align}
    \Phi_{\varepsilon,i} &= \int \limits_{p_i = p(t_0 + \Delta t)}^{p_u = p(t_0)} 4 \pi p^2 T(p) f(p,t_0) \dd{p} \,,
    \label{eq: CR energy flux}
\end{align}
where we have the same upper integration boundary $p_u$ as before. Again, we use the general expression~\eqref{eq: relativistic (kinetic) energy} for the kinetic energy and the case distinction~\eqref{eq: case distinction for distribution function}:
\begin{equation*}
\begin{aligned}
	\Phi_{\varepsilon,i} 
      &= \begin{cases}
       4 \pi c (mc)^4 f_i \hat{p}_i^{q_i} \int \limits_{\hat{p}_i}^{\hat{p}_u} \hat{p}^{2-q_i}\, \left(\sqrt{\hat{p}^2 + 1} -1 \right) \dd{\hat{p}},  &\hat{p}_u > \hat{p}_i \,,\\
       4 \pi c (mc)^4 f_{i-1} \hat{p}_{i-1}^{q_{i-1}} \int \limits_{\hat{p}_i}^{\hat{p}_u} \hat{p}^{2-q_{i-1}} \left(\sqrt{\hat{p}^2 + 1} -1 \right) \dd{\hat{p}}\,, \!\!  &\hat{p}_u < \hat{p}_i\,.
   \end{cases}
\end{aligned}
\end{equation*}

\subsubsection{Criterion for the simulation timestep}\label{sec: Criterion for the simulation timestep}
 
In order to compute the fluxes $\Phi_{n,i}$ and $\Phi_{\varepsilon,i}$ in the $i$-th bin reliably, the timestep of an arbitrary momentum loss process must be small enough so that $p_u$ is not larger than the upper boundary $p_{i+1}$ of the current bin, because in this extreme case the bin would already be emptied completely in one timestep (an analogous argument can be made for momentum gains). Consequently, we require that $p_u = p_{i+1}$ when $\Delta t = \Delta t_\mathrm{max}$: 
\begin{equation}
    p_u(\Delta t_\mathrm{max}, p_i) = p_{i+1}\,.
    \label{eq: bin-criterion for maximum timestep}
\end{equation}
Since this still depends on the momentum bin, in Sec.~\ref{sec: Energy loss processes} we compute $\Delta t_\mathrm{max}$ for the bin $[p_j, p_{j+1}]$, where it is shortest. In the case of Coulomb and bremsstrahlung losses $\Delta t_\mathrm{max}$ is shortest for the lowest bin and vice versa for synchrotron/IC losses.

After the computation of $\Delta t_\mathrm{max}$ for all loss processes we can split each simulation timestep $\Delta t_\mathrm{sim}$ into $N_\mathrm{sub}= \lceil \Delta t_\mathrm{sim}/ \Delta t_\mathrm{sub} \rceil$ subcycling timesteps, where $\Delta t_\mathrm{sub}$ is set by the shortest loss timescale:
\begin{align}
    \Delta t_\mathrm{sub} = C_\mathrm{Cour}\, \min(\Delta t_{\mathrm{max, Coulomb}}, \Delta t_{\mathrm{max, synch+IC}}, \dots)\,.
    \label{eq: computation of Delta t_sub}
\end{align}
We found that multiplying the minimum timestep with a Courant-like factor $C_\mathrm{Cour}=0.8$ is a good compromise between performance and stability of the CR solver. 

If $N_\mathrm{sub}$ becomes too large, a simulation will slow down significantly. Hence, we restrict the maximum number of subcycles to a default value of $N_\mathrm{sub,max}=8$, which we found to be suitable for most cosmological simulations. We will revisit the constraints of short subcycling timesteps later in Sec.~\ref{sec: Constraints on timesteps}.

\subsection{Momentum loss processes} \label{sec: Energy loss processes}

We distinguish between continuous and catastrophic energy loss mechanisms. The former preserve the total number of CRs and only shift them between neighbouring bins, whereas the latter act like negative source terms, i.e. CRs are destroyed. This is discussed in more detail in App.~\ref{app: Continuous and catastrophic loss processes}. In brief, the general differential equation describing a continuous momentum loss processes $\dot{p} = -b(p)$ is
\begin{align*}
    \pdv{f(p,t)}{t} = \frac{1}{p^2} \pdv{p} (p^2 b(p) f(p,t))\,, \quad f(p,t_0) = f_\mathrm{ini}(p)\,,
\end{align*}
and has the solution
\begin{equation}
    f(p,t) = f_\mathrm{ini}(p_u(p, t)) \frac{p_u^2(p, t)}{p^2} \frac{b(p_u(p,t))}{b(p)}\,,\;\; \int \limits_p^{p_u} \!\! \frac{\dd{s}}{b(s)} = t-t_0\,,
    \label{eq: analytical solution for f due to loss processes}
\end{equation}
where the second equation implicitly defines $p_u$ as in Eq.~\eqref{eq: implicit definition of pu}.

For electrons we consider adiabatic, Coulomb, bremsstrahlung, synchrotron and inverse Compton (IC) losses, where the latter two are combined in one term due to their strong similarity:
\begin{align*}
    {b}_{\mathrm{tot,e}}(\hat{p})
	= {b}_{\mathrm{ad}}(\hat{p}) + b_\mathrm{Coul}(\hat{p}) + b_\mathrm{brems}(\hat{p}) +{b}_{\mathrm{syn+IC}}(\hat{p}) \,. 
\end{align*}
For convenience, we expressed all terms via the dimensionless momentum $\hat{p}$. On the other hand, for protons we consider adiabatic, Coulomb and hadronic losses:
\begin{align*}
    {b}_{\mathrm{tot,p}}(\hat{p})
	= {b}_{\mathrm{ad}}(\hat{p}) + b_\mathrm{Coul}(\hat{p}) + b_\mathrm{had}(\hat{p}) \,.
\end{align*}
In the following sections we summarize the relevant equations for the loss rate $b(\hat{p})$, initial momentum $\hat{p}_u$, cut-off $\hat{p}_\mathrm{cut}$ and maximum timestep $\Delta t_\mathrm{max}$. More details can be found in App.~\ref{app: Summary of continuous loss processes}.

\subsubsection{Adiabatic changes} \label{sec: Adiabatic changes}

Adiabatic momentum changes of cosmic rays are described by
\begin{align}
    \dot{\hat{p}}_\mathrm{ad} = -b_\mathrm{ad}(\hat{p}) = -B_\mathrm{ad}\, \hat{p} \,, \quad B_\mathrm{ad} =  \frac{1}{3} (\div{\vb{u}})\,.
    \label{eq: adiabatic loss rate}
\end{align}
An adiabatic expansion/compression of the background plasma, where CRs lose/gain momentum, corresponds to a positive/negative divergence of its velocity field $\vb{u}$. This follows directly from the Lagrangian form of the continuity equation:
\begin{equation*}
      \dv{\ln\rho}{t} = - \div{\vb{u}} \,.
\end{equation*}
The analytical solution for $\hat{p}(t)$,
\begin{align}
    \hat{p}(t_0+{\mathrm{\Delta }}t) &= \hat{p}(t_0)\exp (-{B}_\mathrm{ad}{\mathrm{\Delta }}t)\,,
   \label{eq: spectral cut-off for adiabatic losses}
\end{align}
is used to update the spectral cut-off. As discussed after Eq.~\eqref{eq: IVP for b(p)} the initial momentum $p_u$ is obtained by solving \eqref{eq: adiabatic loss rate} with reverse sign:
\begin{align*}
   \hat{p}_{u} &= \hat{p}_i \exp ({B}_\mathrm{ad}{\mathrm{\Delta }}t) \,.
\end{align*}
Last but not least, Eq.~\eqref{eq: bin-criterion for maximum timestep} yields the maximum timestep
\begin{equation}
    \Delta t_\mathrm{max} = \frac{1}{\abs{B_\mathrm{ad}}}\, \ln(\frac{\hat{p}_{i+1}}{\hat{p}_i})\,, 
\end{equation}
which is independent of the bin for constant momentum ratios.

\subsubsection{Coulomb losses} \label{sec: Coulomb losses}
    
    In a fully ionized plasma the momentum loss rate of both CR protons and electrons due to collisions with the plasma electrons is given by \citep[cf.][]{Winner.etal2019, Girichidis.etal2020}
    \begin{align}
        \dot{\hat{p}}_\mathrm{Coul} &= -B_\mathrm{Coul} (\hat{p}^{-2} + 1)\,,  \label{eq: approximation for dp/dt Coulomb}
    \end{align}
    where only the prefactors are different (for details see App.~\ref{app: Summary of continuous loss processes}):
        \begin{align}
        B_\mathrm{Coul,p} &= 0.9\, \frac{3 \sigma_\mathrm{T} n_\mathrm{e} m_\mathrm{e} c}{2 m_\mathrm{p}}\, \ln \left(\frac{2 m_\mathrm{e} c^2}{\hbar \omega_{\mathrm {pl}} }\right)\,, \\
        B_\mathrm{Coul,e} &= \frac{3 \sigma _\mathrm{T} n_\mathrm{e}c}{2}\, \ln \mathopen {} \left(\frac{m_\mathrm{e}c^2}{\hbar \omega _\mathrm{pl}}\right) \,. 
    \end{align}
    Here, $\sigma_\mathrm{T}$ is the Thomson cross section, $n_\mathrm{e}$ is the number density of free plasma electrons and $\omega_\mathrm{pl} = \sqrt{4 \pi e^2 n_\mathrm{e}/ m_\mathrm{e}\, }$ is the plasma frequency. Hence, the initial momentum $p_u$, implicitly defined via Eq.~\eqref{eq: implicit definition of pu}, also has the same form for both particle species (suppressing the subscripts in the prefactor $B_\mathrm{Coul}$):
   \begin{align*}
	\Delta t &= \int \limits_{p_u}^{p_i} \frac{\dd{p}}{\dot{p}} = -\frac{1}{B_\mathrm{Coul}} \int \limits_{\hat{p}_u}^{\hat{p}_i} \frac{\dd{\hat{p}}}{\hat{p}^{-2} + 1} 
    = 
    \left. -\frac{1}{B_\mathrm{Coul}}\left[ \hat{p} - \arctan (\hat{p}) \right] \right|_{\hat{p}_u}^{\hat{p}_i} .
\end{align*}
Formally, this integral leads to an implicit equation for $\hat{p}_u$,
\begin{equation}
    \hat{p}_u - \arctan(\hat{p}_u) = \hat{p}_i - \arctan(\hat{p}_i) + B_\mathrm{Coul} \Delta t\,,
    \label{eq: pu Coulomb from implicit equation}
\end{equation}
which is fully consistent with the loss rate~\eqref{eq: approximation for dp/dt Coulomb} and therefore allows a reliable reconstruction of the spectral slopes after CR numbers and energies have been updated (see App.~\ref{app: Computation of the initial momentum for CR fluxes}). On the other hand, \citet{Winner.etal2019} proposed an explicit solution (with poorer slope reconstruction) by using the approximation
\begin{align*}
    \hat{p} - \arctan (\hat{p}) \approx \frac{\hat{p}^3}{3 + \hat{p}^2}\,,
\end{align*}
which leads to a cubic equation for $\hat{p}_u$,
\begin{align}
   \frac{\hat{p}_u^3}{3 + \hat{p}_u^2} &= \frac{\hat{p}_i^3}{3 + \hat{p}_i^2} + B_\mathrm{Coul}\Delta t\,,
   \label{eq: cubic equation for pu Coulomb}
\end{align}
with one real solution:
\begin{equation} 
    \begin{aligned}
\hat{p}_u(\hat{p}_i, \Delta t) &= \frac{1}{3} \left[a + \left(a^3 + \frac{9}{2} \sqrt{4a^4 + 81 a^2} + \frac{81 a}{2} \right)^{1/3} \right.\\
 & \left. +\: a^2 \left(a^3 + \frac{9}{2} \sqrt{4a^4 + 81 a^2} + \frac{81 a}{2} \right)^{-1/3} \right], \\
 a &\coloneqq \hat{p}_i^3 / (3 + \hat{p}_i^2) + B_\mathrm{Coul} \Delta t\,.
    \end{aligned}
    \label{eq: pu Coulomb Winner}
\end{equation}
Nonetheless, from Eq.~\eqref{eq: cubic equation for pu Coulomb} we obtain a good estimate for $\Delta t_\mathrm{max}$:
\begin{equation}
    \Delta t_\mathrm{max}(\hat{p}_i, \hat{p}_{i+1}) = \frac{1}{B_\mathrm{Coul}} \left( \frac{\hat{p}_{i+1}^3}{3 + \hat{p}_{i+1}^2} - \frac{\hat{p}_i^3}{3 + \hat{p}_i^2}\right) \,.
    \label{eq: Delta tmax Coulomb}
\end{equation}
Since the ratio $\hat{p}_{i+1}/\hat{p}_i \eqqcolon A$ is identical for all bins, we can rewrite $\Delta t_\mathrm{max}$ as a function of $\hat{p}_i$ alone. A quick inspection shows that $\Delta t_\mathrm{max}(\hat{p}_i, A \hat{p}_i)$ is a strictly monotonically increasing function of $\hat{p}_i$, therefore the maximum timestep is shortest for the lowest bin with $\hat{p}_i = \hat{p}_\mathrm{min}$. Note that we do not update the spectral cut-off, because Coulomb losses primarily affect the low-momentum end of the CR spectrum.

\subsubsection{Synchrotron and inverse Compton losses}

Synchrotron and inverse Compton losses were already implemented and tested in our previous work \citep{Boess.etal2023}, so we only summarize the main points here. The combined loss rate
\begin{equation}
    \dot{\hat{p}} = - B_\mathrm{syn+IC} \hat{p}^2\,, \quad B_\mathrm{syn+IC} = \frac{4}{3} \frac{{\sigma }_\mathrm{T}}{m_\mathrm{e} c}({u}_{B}+{u}_{\mathrm{rad}})
    \label{eq: dp/dt for synch+IC}
\end{equation}
depends on the energy densities $u_B$ and $u_\mathrm{rad}$ of the magnetic and the radiation field and is valid for ultra-relativistic electrons with $\beta \rightarrow1$ and low photon energies $E_\mathrm{phot}$ with $\gamma_\mathrm{e} E_\mathrm{phot} \ll m_\mathrm{e}c^2$ far away from the Klein-Nishina limit \citep[e.g.][]{Longair2011}. We update the cut-off using the analytical solution of Eq.~\eqref{eq: dp/dt for synch+IC}:
\begin{align*}
    \hat{p}(t) = \frac{\hat{p}_0}{1 + B_\mathrm{syn+IC} \hat{p}_0 t}\,,
\end{align*}
From this we directly obtain the initial momentum,
\begin{align*}
   \hat{p}_u(\hat{p}_i, \Delta t) = \frac{\hat{p}_i}{1 - B_\mathrm{syn+IC} \hat{p}_i \Delta t}\,,
\end{align*}
and the maximum timestep,
\begin{align*}
    \Delta t_\mathrm{max} = \frac{1}{B_\mathrm{syn+IC} \hat{p}_i} \left(1 - \frac{\hat{p}_i}{\hat{p}_{i+1}} \right) \,,
\end{align*}
which is smallest for $\hat{p}_{i+1} = \hat{p}_\mathrm{cut}$ for a fixed ratio $\hat{p}_i/\hat{p}_{i+1}$.

\subsubsection{Bremsstrahlung losses}

In a fully ionized H-He plasma with helium mass fraction $Y$, the momentum loss rate of a CR electron due to electron-electron and electron-nucleus bremsstrahlung is
\begin{equation}    
    \dot{\hat{p}} = -B_\mathrm{brems} (\hat{p} + \hat{p}^{-1}) \,, \quad 
    B_\mathrm{brems} =  \frac{3 \alpha _\mathrm{ fs} \sigma _\mathrm{ T} c \bar{g}}{2 \pi} n_\mathrm{e} \frac{4 -Y}{2-Y} \,, 
\label{eq: approximation for dp/dt bremsstrahlung electrons}
\end{equation} 
where $\alpha_\mathrm{fs}$ is the fine structure constant, $n_e$ is the number density of plasma electrons, and for the Gaunt factor we use a constant value of $\bar{g} \approx 5.7$ (c.f. App.~\ref{app: Summary of continuous loss processes}).

By inserting the loss rate~\eqref{eq: approximation for dp/dt bremsstrahlung electrons} into \eqref{eq: implicit definition of pu},
   \begin{align*}
	\Delta t = \int \limits_{p_u}^{p_i} \frac{\dd{p}}{\dot{p}} = -\frac{1}{B_\mathrm{brems}} \int \limits_{\hat{p}_u}^{\hat{p}_i} \frac{\dd{\hat{p}}}{\hat{p} + \hat{p}^{-1}}  = -
    \left. \frac{1}{2 B_\mathrm{brems}} \ln(1+\hat{p}^2) \right|_{\hat{p}_u}^{\hat{p}_i} \,,
\end{align*}
we get an explicit expression for $\hat{p}_u$,
    \begin{align*}
\hat{p}_u &= \sqrt{ (1+\hat{p}_i^2)\, \exp(2B_\mathrm{brems} \Delta t) -1} \,,
    \end{align*}
and for the maximum timestep~\eqref{eq: bin-criterion for maximum timestep},
    \begin{align*}
    \Delta t_\mathrm{max} = \frac{1}{2B_\mathrm{brems}} \ln(\frac{1+\hat{p}_{i+1}^2}{1+\hat{p}_i^2}) \,, 
    \end{align*}
where $\Delta t_\mathrm{max}$ is again shortest for the lowest momenta bin with $\hat{p}_i = \hat{p}_\mathrm{min}$. The spectral cut-off $\hat{p}_\mathrm{cut}$ can be updated using the ultra-relativistic approximation:
\begin{align*}
    \dot{\hat{p}} \approx -B_\mathrm{brems} \hat{p} \quad \Longrightarrow \quad \hat{p}_\mathrm{cut} = \hat{p}_\mathrm{cut,old} \exp(-B_\mathrm{brems}t) \,.
\end{align*}
If both synchrotron/IC and bremsstrahlung losses are activated in a simulation then we use the analytical solution
\begin{align*}
    \hat{p}_\mathrm{cut} = \frac{B_\mathrm{brems} \hat{p}_\mathrm{cut,old}}{B_\mathrm{brems} \exp(B_\mathrm{brems}t) + B_\mathrm{syn+IC}(\exp(B_\mathrm{brems} t)-1) \hat{p}_\mathrm{cut,old}}
\end{align*}
for the combined loss rate of $\dot{\hat{p}} \approx -B_\mathrm{brems} \hat{p} - B_\mathrm{syn+IC}\hat{p}^2$.

\subsubsection{Hadronic losses}\label{sec: hadronic losses}

A CR proton $p^+$ colliding with another proton from the background plasma can produce a neutral pion $\pi^0$ in the reaction
\[ p^+ + p^+ \rightarrow  p^+ + p^+ + \pi^0  \]
if its momentum is larger than the threshold momentum $p_\mathrm{thr}=0.78\;\si{GeV/c} = 0.828~m_\mathrm{p} c$.
For charged pions $\pi^+$ and $\pi^-$ the threshold momentum is larger, because (1) their masses are slightly higher and (2) the most important production channels,
\begin{align*}
    p^+ + p^+ &\rightarrow p^+ + n + \pi^+ \,,
    \qquad p^+ + p^+ \rightarrow p^+ + p^+ + \pi^+ + \pi^-\,,
\end{align*}
are energetically less favourable \citep{Owen.etal2023}.
The net effect is that in one inelastic collision the CR proton suddenly loses a significant fraction $K_\mathrm{p} \approx 0.5$ of its initial kinetic energy, so this loss process is not continuous. On the other hand, it is also not purely catastrophic, because the original proton is not destroyed in the reaction (except when a $p \rightarrow n$ conversion occurs). Nevertheless, both approximations have been used in the literature and we will discuss the catastrophic approach in App.~\ref{app: Summary of continuous loss processes}. For now, we focus on the continuous approximation, where the energy loss rate is given by $\mathrm{d}E_\mathrm{kin}/\mathrm{d}t = - \nu_\mathrm{pp} \Delta E_\mathrm{kin}$ with the frequency $\nu_\mathrm{pp}$ for inelastic proton-proton collisions and the typical amount of energy $\Delta E_\mathrm{kin} = K_\mathrm{p} E_\mathrm{kin}$ that is lost during such an interaction, where $K_\mathrm{p} \approx 0.5$. The momentum loss is then given by \citep{Mannheim.Schlickeiser1994, Ensslin.etal2007, Girichidis.etal2020, Werhahn.etal2021}:
    \begin{equation}
    \begin{aligned}
         \dot{\hat{p}}_\mathrm{had} &= - B_\mathrm{had} \left(\sqrt{\hat{p}^2 +1} -1 \right)H(\hat{p} - \hat{p}_\mathrm{thr}) \,, \\
         B_\mathrm{had} &=  n_\mathrm{N} \sigma_\mathrm{pp} K_\mathrm{p} c \,.
    \end{aligned}
    \label{eq: hadronic losses: continuous approach}
    \end{equation}
Here, the Heaviside function ensures that only for $\hat{p} > \hat{p}_\mathrm{thr}$ the loss rate is non-zero, $n_\mathrm{N} = n_\mathrm{e}/(1-Y/2) = n_\mathrm{H} + 4 n_\mathrm{He} \approx 1.3~n_\mathrm{H}$ is the density of target nucleons in a pure H-He-plasma with helium mass fraction $Y$, and $\sigma_\mathrm{pp} \approx 30$~mbarn is the total inelastic cross section for pion production, which is often assumed to be constant. Although more detailed formalisms with an analytical parametrizations of the energy-dependent cross section exist \citep[e.g.][]{Krakau.Schlickeiser2015}, the error of this simplification only grows logarithmically with particle energy, as pointed out by \citet{Mannheim.Schlickeiser1994}. 

In a bin below the threshold momentum no hadronic losses can take place, so for the initial momentum $\hat{p}_u$ we simply get $\hat{p}_u= \hat{p}_i$ for $\hat{p}_i < \hat{p}_\mathrm{thr}$. For the case $\hat{p}_i \geq \hat{p}_\mathrm{thr}$ Eq.~\eqref{eq: implicit definition of pu} gives an implicit equation for $\hat{p}_u$: 
\begin{align}
    B_\mathrm{had} \Delta t = \! \int \limits_{\hat{p}_i}^{\hat{p}_u} \! \frac{\dd{\hat{p}}}{\sqrt{1 + \hat{p}^2}-1} = \left. \frac{\hat{p}\,  \mathrm{arsinh}(\hat{p}) - 1 - \sqrt{1+ \hat{p}^2} }{\hat{p}} \right|_{\hat{p}_i}^{\hat{p}_u} \!.
    \label{eq: pu hadronic implicit}
\end{align}
As for Coulomb losses this is the most self-consistent way of computing $\hat{p}_u$, but at the cost of an iterative method.

Although no explicit solution for $\hat{p}_u$ exists, there are approximations for both the non- and ultra-relativistic regime:
\begin{equation}
        \begin{aligned}
        &\text{non-rel.:} & \!\!\!\! \dot{\hat{p}} &= -B_\mathrm{had} \hat{p}^2/2 
        &\Longrightarrow\; \hat{p}_{u, \mathrm{n.r}}(\hat{p},t) &= \frac{2 \hat{p}}{2 - B_\mathrm{had} \hat{p} t}\,, \\
        &\text{ultra-rel.:} &\!\!\!\! \dot{\hat{p}} &= -B_\mathrm{had} \hat{p}
        &\Longrightarrow\; \hat{p}_{u, \mathrm{u.r}} (\hat{p} ,t) &= \hat{p} \exp(B_\mathrm{had} t)\,.
    \end{aligned}
    \label{eq: hadronic losses non-rel and ultra-rel limit}
\end{equation}
The ultra-relativistic approximation can be conveniently used for the timestep constraint, because for $\hat{p} \lesssim 1$ it is shorter (and thus more restrictive) than the proper estimate from the first line in Eq.~\eqref{eq: hadronic losses non-rel and ultra-rel limit}. The result is independent of $\hat{p}_i$ for a constant ratio $\hat{p}_{i+1}/\hat{p}_i$:
    \begin{align*}
        \Delta t_\mathrm{max} = \frac{1}{B_\mathrm{had}} \ln(\frac{\hat{p}_{i+1}}{\hat{p}_i})\,.
    \end{align*}
Similarly, we update the cut-off using the ultra-relativistic simplification $\hat{p}_\mathrm{cut}(t_0+\Delta t) = \hat{p}_\mathrm{cut}(t_0) \exp(- B_\mathrm{had} \Delta t)\,.$

\subsection{Seeding of cosmic rays by supernova remnants}

The default subgrid model for the multiphase interstellar medium in \textsc{OpenGadget3} in the effective model by \citet{Springel.Hernquist2003}, where a gas particle above a critical threshold density begins to ``form stars'', meaning that it injects thermal energy into the ambient medium corresponding to heating by core-collapse supernovae that occur shortly after star formation has begun. Besides thermal energy, also magnetic fields and cosmic rays can be injected until the gas particle will eventually be converted into a star particle, which represents a stellar population characterized by an initial-mass function.

\subsubsection{Injection of power-law spectra}\label{sec: Injection of power-law spectra}
To illustrate our subgrid description for cosmic-ray injection by supernova remnants we consider an SPH gas particle with ongoing supernova activity. Some distribution function $f(\vb{x},p)$ represents the total number of CRs that a supernova remnant produces over its lifetime and eventually releases into the surrounding ISM. The ``CR spectrum'' $N(p)$ associated with the corresponding SPH particle is the integral of $f(\vb{x},t)$ over a volume large enough so that it contains all newly produced CRs, which we assume to be the SPH particle's volume $V_\mathrm{SPH}$, i.e. a ball with a radius equal to the hydrodynamical smoothing length:
\begin{align*}
    N(p) = \int_{V_\mathrm{SPH}} f(\vb{x}, p) \dd{\vb{x}}\,.
\end{align*} 
If $N_\mathrm{SN}$ supernova explosions occur in the gas particle during one timestep, then, to the lowest order, the freshly injected CR population is just the superposition $N_\mathrm{SN} \cdot N(p)$, which is numerically represented by the arrays of CR numbers and energies per unit mass. They are computed by replacing $f(p)$ with $N(p)$ in Eqs.~\eqref{eq: CR number integral} and \eqref{eq: CR energy integral} and dividing through the mass of the SPH particle instead of its density. This subgrid model is valid if the following two conditions are met. First, the simulation timestep should be larger than the timespan over which a supernova remnant can accelerate CRs. One typically assumes that CR acceleration lasts until the end of the Sedov-Taylor phase \citep[see the discussion in][and references therein]{Morlino.Celli2021}, which is reached after a few $10^4$~years. Second, all newly produced CRs should be enclosed by the volume of the SPH particle, so the hydrodynamical smoothing length should be larger than the distance over which the CRs can propagate in the aforementioned timespan. 

In the simplest version of this subgrid model one assumes that the injected spectrum is a power-law in momentum space \citep{Ensslin.etal2007, Jubelgas.etal2008, Werhahn.etal2021, Girichidis.etal2022, Hopkins.etal2022a}:
\begin{equation*}
    N(p) = \begin{cases}
        K p^{-q} \quad &\mathrm{if}\; p_\mathrm{min} \leq p \leq p_\mathrm{max}\,, \\
        0 \quad &\mathrm{otherwise}\,.
    \end{cases}
\end{equation*}
It is characterized by four parameters: the spectral slope $q$, the normalization $K$, and the cut-offs $p_\mathrm{min}$ and $p_\mathrm{max}$ at the low- and high-energy end of the spectrum. We will briefly outline how each of these parameters can be motivated.

The test-particle theory of diffusive shock acceleration predicts $q=4$ for strong shocks \citep{Blandford.Eichler1987}, but recent particle-in-cell (PIC) simulations have shown that the spectral slope should be slightly steeper, i.e. $q\gtrsim 4.2$ (in agreement with observations of the Galactic CR spectrum). The main reason is that downstream MHD waves scattering CRs are isotropized in a postcursor region moving away from the shock faster than the gas advection speed. This increases the probability that cosmic rays escape from the acceleration region around the shock, which leads to a steeper spectrum \citep{Caprioli.etal2020, Diesing.Caprioli2021, Cristofari.etal2022}. Since electrons have the same rigidity as protons their slope mainly differs for high momenta, where efficient radiative loss processes steepen the electron spectrum. 

The spectrum is usually normalized so that the integrated kinetic energy of CR protons is ${\sim} 10\%$ of the typical kinetic energy $E_\mathrm{SN} \approx 10^{51}$~erg of the supernova ejecta, which can vary by more than one order of magnitude amongst different supernovae \citep{Martinez.etal2022}. This normalization is motivated by comparing the Galactic CR luminosity (CR energy content of the Galactic disk divided by diffusive escape time) to the Galactic supernova luminosity (explosion energy $E_\mathrm{SN}$ times supernova rate). If ${\sim} 10\%$ of the supernova energy was converted into CRs, this would already be sufficient to maintain the observed Galactic CR luminosity \citep[cf.][and references therein]{Drury2012}. Moreover, the CR electron spectrum is usually rescaled by an electron-to-proton ratio $K_\mathrm{ep}$. \citet{Schlickeiser2002} showed that by assuming power-law distribution functions for both electrons, $f_e(p)=f_{e,0} p^{-q}$, and protons, $f_p(p)= f_{p,0} p^{-q}$, starting at the same kinetic energy in the supra-thermal pool, this ratio is
\begin{align*}
	K_\mathrm{ep} \coloneqq \frac{f_\mathrm{e}(p)}{f_\mathrm{p}(p)} = \frac{f_{0,e}}{f_{0,p}} \approx \quantity(\frac{m_e}{m_p})^{(q-3)/2} \,.
\end{align*}
For $q=4.2$ this yields $K_\mathrm{ep} \approx 0.01$, which agrees well with the observed ratio in the local ISM \citep[for a detailed discussion see][]{Merten.etal2017}. 

The high-momentum cut-off $p_\mathrm{max}$ is motivated by the steepening of the total Galactic CR spectrum at energies around ${\sim}3$~PeV, which is called the ``knee''. Therefore, Galactic CR sources should be able to accelerate CR protons up to PeV energies \citep{Gabici.etal2019}. However, this is hard to achieve for normal supernova remnants, as shown by multiple studies \citep[e.g.][]{Cristofari.etal2020, Inoue.etal2021, Diesing2023, Brose.etal2025}, so assuming a spectral cut-off $\hat{p} \lesssim 10^6$ for protons is reasonable. On the other hand, electrons with $\hat{p}\gtrsim 10^6$ have very short cooling times and barely contribute to the integrated energy.

The low-momentum cut-off of CR protons is usually set to $\hat{p}_\mathrm{min,p} \gtrsim 10^{-2}$, because Coulomb and ionization losses are very efficient in this regime. Moreover, at sub-GeV energies the transition from the CR population to the the thermal pool starts, which is not captured by our model. We emphasize that the total energy budget of a typical CR proton spectrum is dominated by particles with $\hat{p} \in [1,10]$. Hence, as long as $\hat{p}_\mathrm{min,p}$ and $\hat{p}_\mathrm{max,p}$ are not too close to this interval, the spectral normalization does not change significantly if the integrated CR energy is fixed. 
The cooling time of CR electrons is longest for $10^2 \lesssim \hat{p} \lesssim 10^3$ and becomes extremely short for $\hat{p} \lesssim1$ \citep{Sarazin1999}. These low-momentum electrons neither carry a majority of the integrated energy nor are they relevant for non-thermal radiation signatures, so one can safely set $\hat{p}_\mathrm{min,e} \gtrsim 10$.

\subsubsection{Injection of template spectra}

Modelling the CR spectra released by supernova remnants is challenging due to the large number of physical processes involved. The most important ingredients are (1) the temporal evolution of the shock's radius and speed; (2) the density profile of the ambient medium (relevant especially for core-collapse SNe, where it has been affected by the wind of the progenitor star); (3) the maximum momentum to which CRs can be accelerated at a given time (relying on models for magnetic field amplification and diffusion at the shock front); (4) time-dependent CR injection spectrum and acceleration efficiency at the shock front; (5) energy loss processes for CR protons and electrons. Recent studies have incorporated these items semi-analytically \citep[e.g.][]{Morlino.Celli2021, Cristofari.etal2020, Cristofari.etal2021, Diesing.Caprioli2021} or by numerically solving the hydrodynamical equations for the remnant's evolution and the CR transport equation \citep[e.g.][]{Zirakashvili.Ptuskin2012, Das.etal2022, Das.etal2024}.

\begin{figure}
    \centering
    \includegraphics[width=\linewidth]{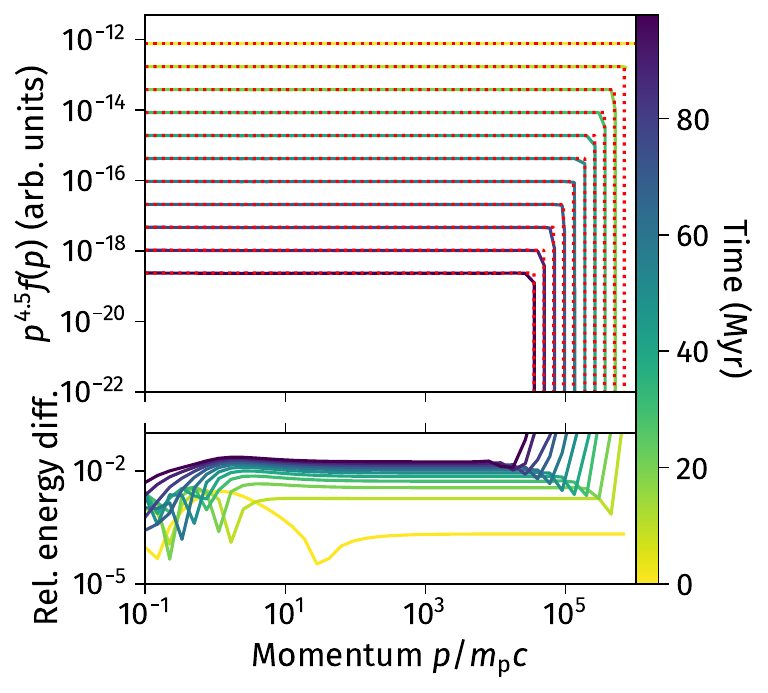}
    \caption{Cooling of a CR proton spectrum in an adiabatically expanding background medium with $\div{\vb{u}}=0.1~\mathrm{Myr^{-1}}$, where the analytical solution~\eqref{eq: analytical solution for adiabatic losses} is overplotted in red. The bottom panel shows the relative energy error from Eq.~\eqref{eq: relative energy error} for each bin.}
    \label{fig: CRp adiabatic expansion test}
\end{figure}

The big advantage of these more sophisticated models is that the total CR energy budget, spectral slope and the (less relevant) cut-offs are not inserted by hand, but result from physically motivated input parameters. Despite their quantitative differences, all models come to the main conclusion that the overall spectrum of cosmic rays, that are released into the ISM by a supernova remnant, consists of two parts, namely 1) a high-energy tail of CRs that cannot be confined at the shock front anymore and escape upstream and 2) lower energy CRs that are advected downstream after leaving the acceleration cycle at the shock and are confined within the remnant. They eventually escape when the shock becomes weak enough, but will lose some of their energy in the meantime due to adiabatic and radiative (for electrons) losses. This was already found by much earlier studies \citep[e.g.][]{Ptuskin.Zirakashvili2005, Caprioli.etal2010}.

Concrete spectra that we tabulated for production runs are those published (and provided) by \citet{Cristofari.etal2021}, which were computed for 3 types of SNe, namely type~Ia, default core-collapse and very energetic core-collapse with $E_\mathrm{SN}= 10^{52}$~erg. They assumed that the freshly injected CR proton spectrum has a spectral slope $q= 4.3$ and an acceleration efficiency of $0.05$, which is the ratio between CR pressure and upstream ram pressure at the shock. For the case of type Ia supernovae, \citet{Morlino.Celli2021} used a slightly different approach to describe CR escape and energy losses and discuss the resulting differences between their spectra and \citet{Cristofari.etal2021}. We take another set of spectra from \citet{Das.etal2024}, where they focus on core-collapse supernovae for progenitor stars with $20~\msol$ and $60~\msol$ and use sophisticated models for the density profile of the ambient medium, which is shaped by the progenitor star's winds. The corresponding CR spectra for type~Ia supernovae were published by \citet{Brose.etal2020, Brose.etal2021}.

To use these detailed models of CR spectra in our code, we tabulate them in text files and then discretize them to the desired number of bins at the beginning of a simulation. We emphasize that this numerical framework is quite flexible, since the files containing the template CR spectra can be updated easily when more sophisticated models become available. Moreover, these spectra can be coupled to the supernova feedback of any effective star formation model, like the more advanced one presented by \citet{Tornatore.etal2004, Tornatore.etal2007}, which also includes supernova of type Ia. In principle, one could create a collection of template spectra, each computed with slightly different model parameters, and then use any environmental property known during runtime, like supernova type, density and pressure of the ambient ISM, to choose the best fitting spectrum that should be injected. Current cosmological simulation codes with spectrally resolved CRs do not make use of the many spectra available in the literature. The subgrid model proposed here is a first step in this direction.

\begin{figure}
    \centering
    \includegraphics[width=\linewidth]{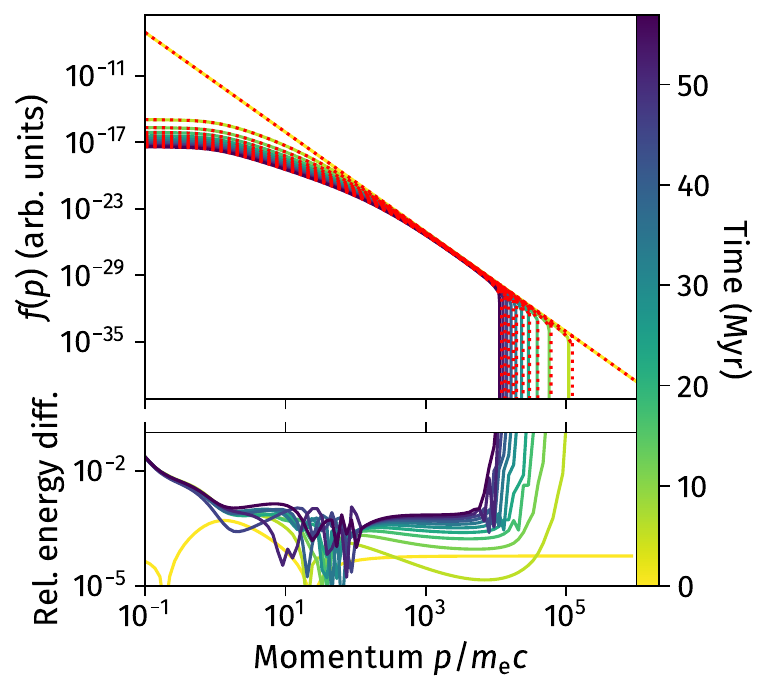}
    \caption{Freely cooling electron spectrum affected by Coulomb, bremsstrahlung and synchrotron/IC losses. The red dotted lines are the analytical solution~\eqref{eq: analytical solution for f due to loss processes}, where $p_u$ was computed for all three loss processes using an ODE solver. Note that on this timescale the effect of bremsstrahlung losses is not significant.}
    \label{fig: all losses for CRe}
\end{figure}

\section{Numerical tests}\label{Sec: Numerical tests}

In this section we show two types of tests to validate the newly implemented features, namely (1) simple ``single particle'' tests to verify that the evolved spectrum agrees with the analytical solution and (2) idealized setups with multiple gas particles to demonstrate that our CR solver also works reliably in a more realistic environment. 

\begin{figure}
    \centering
    \includegraphics[width=\linewidth]{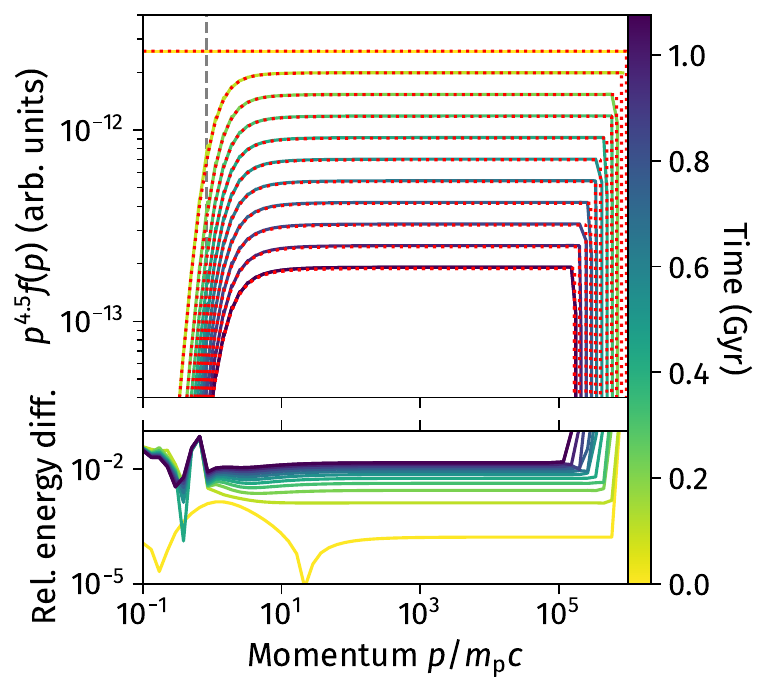}
    \caption{Evolution of CR proton spectrum due to Coulomb~\eqref{eq: approximation for dp/dt Coulomb} and hadronic losses~\eqref{eq: hadronic losses: continuous approach}. The gray dashed line is the threshold momentum $\hat{p}_\mathrm{thr}$, and the red dotted lines are the analytical solution.}
    \label{fig: Coulomb and hadronic losses for protons}
\end{figure}

\subsection{Tests with a single cosmic-ray spectrum}

We start with an initial power-law spectrum,
\begin{equation}
    f(p,t=0) = f_\mathrm{ini}(p) = \begin{cases}
        f_0\, (p/p_\mathrm{min})^{-q_\mathrm{ini}}\,, \!\!\! &p_\mathrm{min} \leq p \leq p_\mathrm{max}\,,\\
        0 & \mathrm{otherwise}\,,
    \end{cases}
    \label{eq: power-law initial condition}
\end{equation}
and update it for each simulation timestep, which we compute according to Eq.~\eqref{eq: computation of Delta t_sub} from the cooling times of all loss processes under consideration. We evolve the spectrum for a total duration of $10^3{-}10^5$ cooling times to demonstrate the core physical effects and the numerical stability of our code. To better illustrate spectral features, like cut-offs or changes in slope, we choose a high resolution with ${\sim}10$~bins per dex in momentum. If not stated otherwise, we compute the different momentum loss rates by assuming a background magnetic field with strength $B=3~\mathrm{\mu G}$, a radiation field dominated by the cosmic microwave background with energy density $u_\mathrm{CMB} = 0.26~\mathrm{eV\,cm^{-3}}$ and an electron number density $n_\mathrm{e} = 10^{-1}~\mathrm{cm^{-3}}$. Motivated by the discussion in Sec.~\ref{sec: CR seeding by supernova remnats}, we choose $q_\mathrm{ini}=4.5$ as the default value for the spectral slope. This is slightly steeper than $q_\mathrm{ini}=4$, where every logarithmic momentum bin would contain the same energy in the ultra-relativistic limit. Since in cooling tests we are mainly interested in the evolution of spectral features, the normalization $f_0$ is unimportant, and $f$ is plotted in arbitrary code units.

For each test, we plot both the numerical solution of the distribution function $f(p,t)$ for different times (coloured lines) and the analytical solution (red dotted lines), which is derived in the corresponding subsections. Due to the high number of bins, the discretization error is small, and both graphs nearly coincide. For a quantitative comparison, we also plot the relative error $\Delta \varepsilon_\mathrm{rel,i}$ of the CR energies in each bin, which is defined as
\begin{equation}
    \Delta \varepsilon_{\mathrm{rel},i} = \abs{1 - \varepsilon_{\mathrm{numerical},i} / {\varepsilon_{\mathrm{analytical},i}}} \,.
    \label{eq: relative energy error}
\end{equation}
Here, $\varepsilon_\mathrm{numerical,i}$ is the CR energy in the $i$-th bin from the simulation output and $\varepsilon_\mathrm{analytical,i}$ is computed by integrating the analytical solution of $f(p,t)$ over the interval $[p_i,p_{i+1}]$.

\begin{figure}
    \centering
    \includegraphics[width=\linewidth]{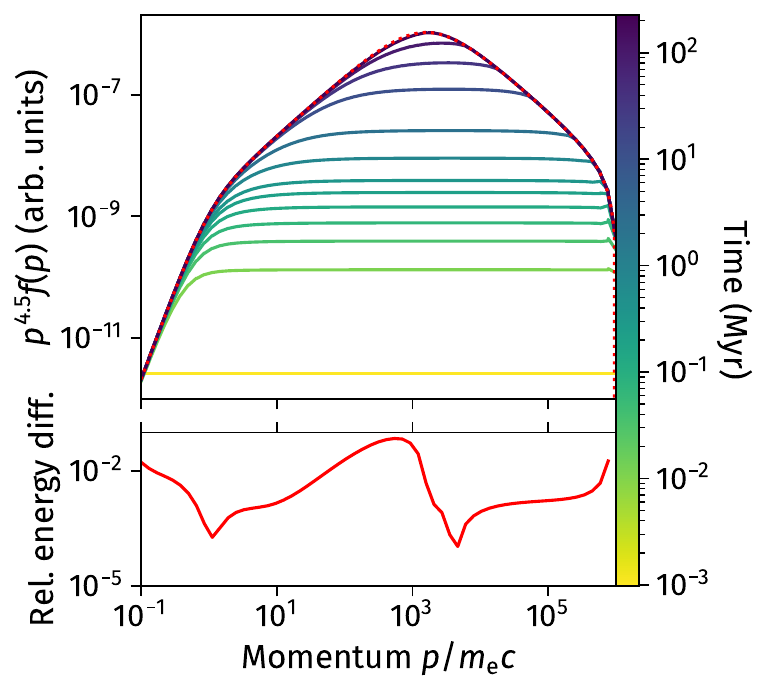}
    \caption{Continuous injection of an electron spectrum at a rate~\eqref{eq: j_inj(p)} that is affected by Coulomb, bremsstrahlung, and synchrotron/IC losses. It clearly converges to the steady-state solution~\eqref{eq: steady-state solution} with $b(p)=b_\mathrm{Coul}(p) + b_\mathrm{brems}(p) + b_\mathrm{syn+IC}(p)$ (red dotted line). The relative energy error is only shown for the latest time.}
    \label{fig: CRe_steady_state}
\end{figure}

\subsubsection{Adiabatic changes}

Consider an adiabatically expanding background plasma, where $\div{\vb{u}} > 0$ is time-independent. Then the CR transport equation~\eqref{eq: CR transport equation} simplifies to
\begin{align*}
	\pdv{f(p, t)}{t} - \left(\frac{1}{3} \div{\mathbf{u}}\right) p \frac{\partial f(p, t)}{\partial p} = 0\,, \qquad f(p,0) = f_\mathrm{ini}(p)
\end{align*}
with the solution
\begin{align}
    f(p,t) = f_\mathrm{ini}\left(p \cdot \exp\left( \frac{1}{3}(\div{\vb{u}})\, t\right) \right)\,.
    \label{eq: analytical solution for adiabatic losses}
\end{align}
As shown in Fig.~\ref{fig: CRp adiabatic expansion test}, for the power-law initial condition~\eqref{eq: power-law initial condition}, an adiabatic expansion exponentially decreases the spectral normalization over time while preserving the slope $q_\mathrm{ini}$ (and vice versa for an adiabatic compression):
\begin{align}
    f(p,t) = f_0 \exp\left( - \frac{1}{3}(\div{\vb{u}})\, t q_\mathrm{ini}\right) \left( \frac{p}{p_\mathrm{min}} \right)^{-q_\mathrm{ini}}, \; p \in[p_\mathrm{min}, p_\mathrm{cut}(t)]
    \label{eq: analytical solution for adiabatic losses for power-law}
\end{align}
In App.~\ref{app: Importance of updating the spectral cut-off} we briefly discuss the problems arising when the variable cut-off $p_\mathrm{cut}(t)$ is not properly updated according to Eq.~\eqref{eq: spectral cut-off for adiabatic losses}, thus highlighting the advantage of our implementation.

\begin{figure*}
		\centering
		\includegraphics[width=\textwidth]{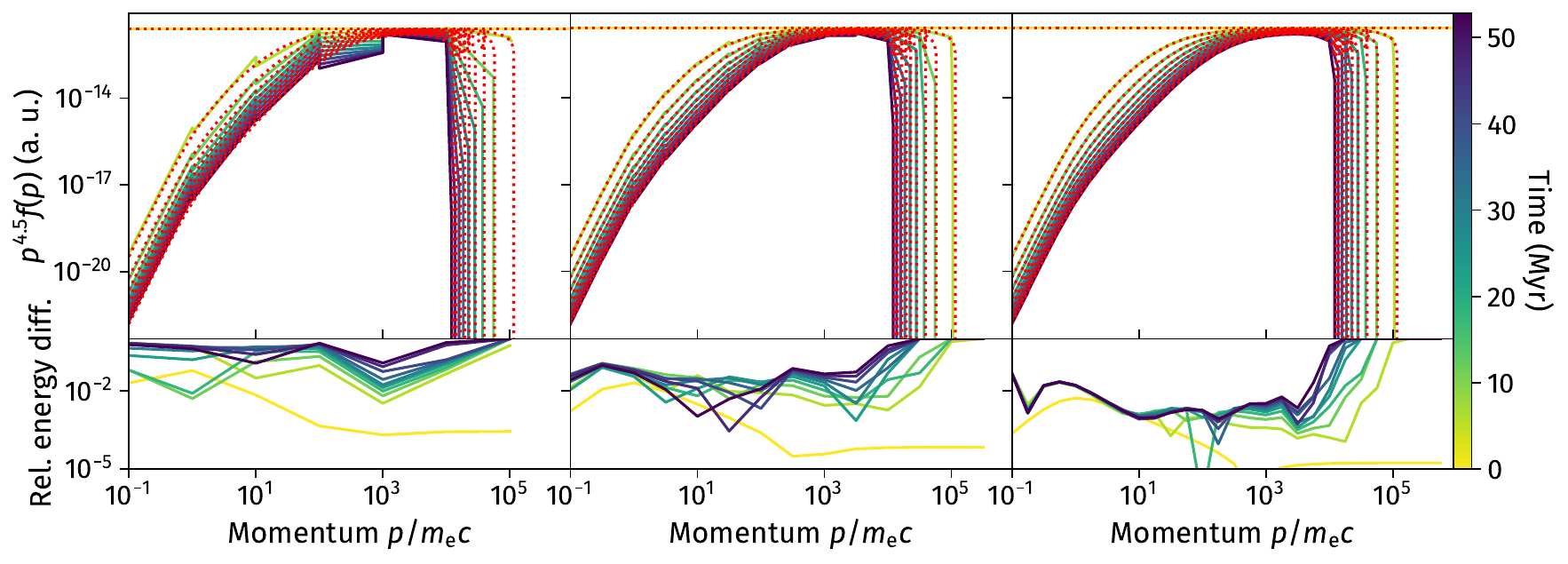}
		\caption{Evolution of a CR electron spectrum cooling under the influence of Coulomb, bremsstrahlung, and synchrotron/IC losses (as in Fig.~\ref{fig: all losses for CRe}) for a duration of $10^5 \cdot \tau_\mathrm{cool} \approx 50$~Myr. The evolved spectrum is shown for different resolutions of 1, 2, and 4 bins/dex (from left to right).}
        \label{fig: CRe_all_losses_different_Nbins}
	\end{figure*}
    
\subsubsection{Freely cooling spectra}\label{sec: Freely cooling spectra}
 
If the differential equation for arbitrary continuous losses,
\begin{align}
    \pdv{f(\hat{p},t)}{t} = \frac{1}{\hat{p}^2} \pdv{\hat{p}} \left( \hat{p}^2 b(\hat{p}) f(\hat{p}, t)\right)\,, \quad f(\hat{p},0)=f_\mathrm{ini}(\hat{p})\,,
    \label{eq: general differential equation for continuous loss process}
\end{align}
cannot be solved explicitly, because the loss rate $b(\hat{p})$ is a complicated function, then Eq.~\eqref{eq: analytical solution for f due to loss processes} provides an analytical solution as long as the weak time-dependence of $b(\hat{p})$ through environmental parameters can be neglected, which we will assume in the following tests. We compute the initial momentum $\hat{p}_u$ by solving the implicit equation~\eqref{eq: implicit definition of pu} using Brent's method, since this results in a more accurate slope reconstruction (see App.~\ref{app: Computation of the initial momentum for CR fluxes}).

Fig.~\ref{fig: all losses for CRe} shows a freely cooling electron spectrum affected by Coulomb, synchrotron/IC and bremsstrahlung losses, where the latter are not important on the timescale considered here. If we include these three loss processes simultaneously, then there is no closed form for $\hat{p}_u$ or the solution~\eqref{eq: analytical solution for f due to loss processes}, which we overplotted in Fig.~\ref{fig: all losses for CRe}. However, since the loss mechanisms are most efficient in different momentum ranges, a qualitative understanding of the full solution can be obtained by considering them individually.

First, Coulomb losses are most relevant in the low-momentum regime with $b_\mathrm{Coul}(\hat{p}) \approx B_\mathrm{Coul} \hat{p}^{-2}$, leading to the analytical solution
\begin{equation*}
    f(\hat{p},t) = f_\mathrm{ini} \left( (\hat{p}^3 + 3B_\mathrm{Coul}t)^{1/3} \right) \,.
\end{equation*}
After a sufficiently long time and for $\hat{p} \lesssim 1$ the term $\hat{p}^3$ becomes negligible and $f(\hat{p},t) \propto \hat{p}^0$ is almost independent of $\hat{p}$, so this part of the spectrum flattens.
Second, bremsstrahlung losses become important for intermediate momenta $\hat{p} \gtrsim 10$ with $b_\mathrm{brems}(\hat{p}) \approx B_\mathrm{brems} \hat{p}$, implying that
\begin{align*}
    f(\hat{p},t) &= \exp(3B_\mathrm{brems}t)\, f_\mathrm{ini} \left( \exp(B_\mathrm{brems}t)\, \hat{p} \right) \\
    &= \exp((3- \alpha_\mathrm{ini})B_\mathrm{brems}t) f_0 (\hat{p}/\hat{p}_\mathrm{min})^{-\alpha_\mathrm{ini}} \,.
\end{align*}
Hence, for $\alpha_\mathrm{ini}>3$ the spectral slope remains constant and the normalization decreases exponentially with time. Third, the simple form of the synchrotron/IC loss rate, $b_\mathrm{syn+IC}(\hat{p}) = B_\mathrm{syn+IC} \hat{p}^2$, allows a fully analytical solution, namely
\begin{equation}
    f(\hat{p},t) = \begin{cases}
        f_\mathrm{ini}\left( \frac{\hat{p}}{1-\hat{p}t B_\mathrm{syn+IC}} \right) \left(1 - \hat{p}t B_\mathrm{syn+IC} \right)^{-4} \hspace{-0.2cm}{,} \hspace{-0.22cm} &\hat{p} <1 /t B_\mathrm{syn+IC}, \\
        0\,, &\hat{p} > 1/t B_\mathrm{syn+IC}.
    \end{cases} 
    \label{eq: Kardashev solution}
\end{equation}
If the initial spectrum $f_\mathrm{ini}(\hat{p})$ is a power-law, as in Eq.~\eqref{eq: power-law initial condition}, then this reduces to the solution presented by \citet{Kardashev1962}, which behaves quantitatively differently for $q_\mathrm{ini}<4$ and $q_\mathrm{ini}>4$. In the latter case the high-momentum end of the spectrum steepens. To better see these effects, we plot $f(p)$ instead of $p^{4.5}f(p)$ in Fig.~\ref{fig: all losses for CRe}.


In Fig.~\ref{fig: Coulomb and hadronic losses for protons} we show the combined effect of Coulomb and (continuous) hadronic losses for CR protons. The former have the same effect as for the electrons, whereas the latter become only relevant for momenta above the threshold $\hat{p}_\mathrm{thr}$ and for $\hat{p} \gtrsim 1$ the loss rate $b_\mathrm{had}(\hat{p}) \approx B_\mathrm{had} \hat{p}$ is similar to an adiabatic expansion. Subsequently, the spectral slope is conserved and the normalization as well as the cut-off decrease exponentially over timescales much larger than the cooling time of electrons.

\subsubsection{Steady-state spectrum}\label{sec: Steady-state spectrum}

\begin{figure*}
    \centering
    \includegraphics[width=0.49 \textwidth]{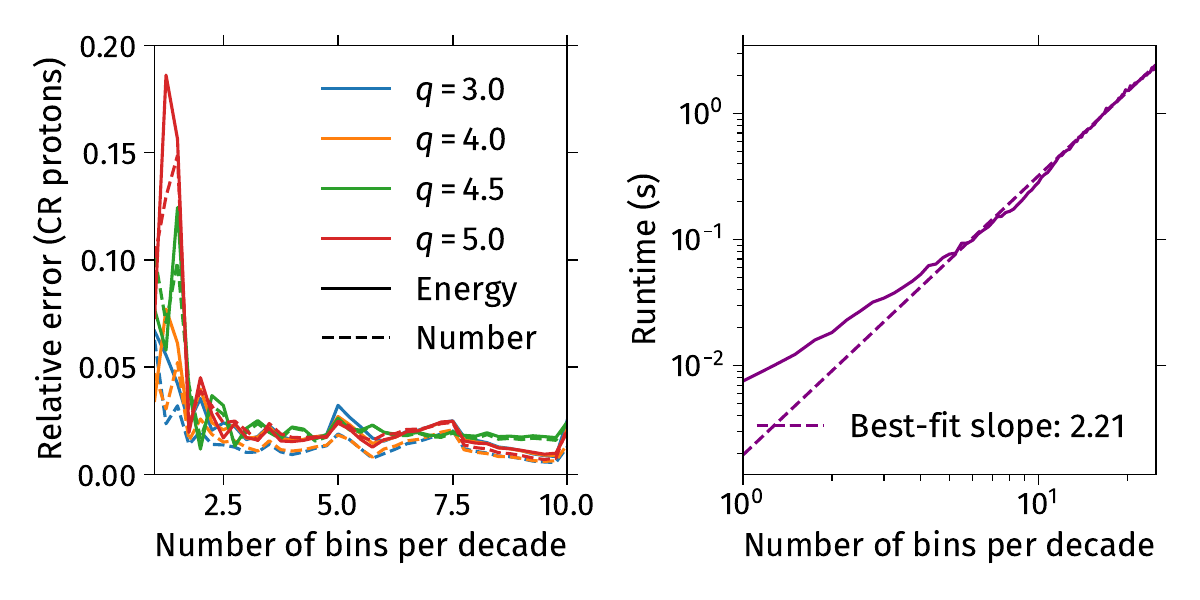}
    \includegraphics[width=0.49 \textwidth]{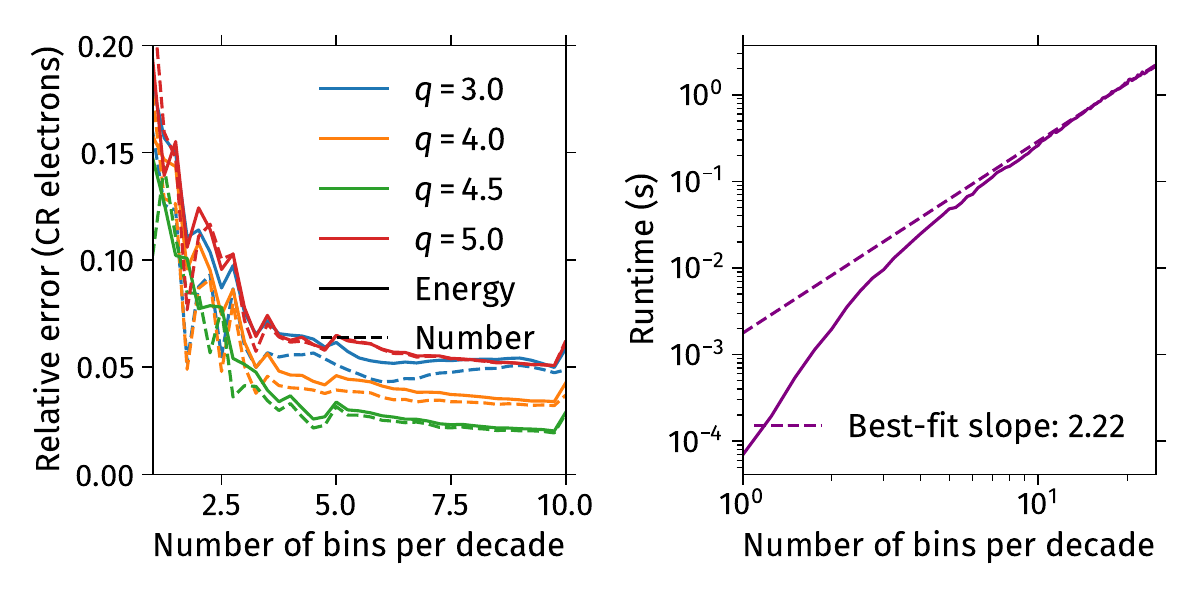}
    \caption{Convergence test for a freely cooling proton spectrum (left, similar to the test in Fig.~\ref{fig: Coulomb and hadronic losses for protons}) and a freely cooling electron spectrum (right, similar to the test in Fig.~\ref{fig: all losses for CRe}). The relative errors for CR numbers (dashed lines) and CR energies (solid lines) are computed as in Eq.~\ref{eq: L1 norm of relative errors} and shown for initial spectra $f_\mathrm{ini}(p) \propto p^{-q}$ with four different slopes (see colours in legends). For the slope $q=4.5$ we also show how the runtime depends on the number of momentum bins, where the dashed line is the best-fit power law for the range $12.5 \leq N_\mathrm{bins}/\mathrm{dex} \leq 25$.}
    \label{fig: convergence test}
\end{figure*}
    
Starting from an initial power-law spectrum~\eqref{eq: power-law initial condition} we continuously inject CR electrons at a rate of 
\begin{equation}
j_\mathrm{inj}(p) = \begin{cases}
    j_0 (p/p_\mathrm{min})^{-q_\mathrm{inj}}\; &\mathrm{if}\; p_\mathrm{min} \leq p \leq p_\mathrm{max}\,,  \\
    0 \; &\mathrm{otherwise}\,.
\end{cases}
\label{eq: j_inj(p)}
\end{equation}
In other words, during each timestep $\Delta t$ the amount of newly added CRs is $j_\mathrm{inj}(p) \Delta t$. By choosing a suitable normalization $j_0 = f_0/(10 \cdot \tau_\mathrm{cool})$ a steady-state spectrum emerges after approximately the maximum cooling time
\begin{align*}
    \max_{p \in [p_\mathrm{min}, p_\mathrm{max}]} \tau(p) = \max_{p \in [p_\mathrm{min}, p_\mathrm{max}]} \abs{p/b(p)}\,, 
\end{align*}
when all momenta of the initial spectrum have been shifted significantly due to loss processes. This is shown in Fig.~\ref{fig: CRe_steady_state} for an electron spectrum, where the continuous injection counterbalances Coulomb, bremsstrahlung and synchrotron/IC losses.

For a homogeneous distribution of sources the CR transport equation for the steady-state spectrum $f(p)$ (without dependence on space and time) is given by
\begin{align*}
    \frac{1}{p^2} \pdv{p} \left( p^2 b(p) f(p)\right) = -j_\mathrm{inj}(p)\,.
\end{align*}
By integrating this from $p$ to $\infty$ with the boundary condition $\lim_{p \rightarrow \infty}f(p)=0$ and assuming that $q_\mathrm{inj} > 3$ one obtains the solution \citep[cf.][]{Longair2011}
\begin{equation}
\begin{aligned}    
    f(p) &= \frac{1}{p^2 b(p)} \int \limits_p^\infty j_\mathrm{inj}(\tilde{p}) \tilde{p}^2 \dd{\tilde{p}} \\
    &= \frac{j_0 p_\mathrm{min}^{q_\mathrm{inj}}}{(q_\mathrm{inj}-3)p^2b(p)} \begin{cases}
        \left( p^{3-q_\mathrm{inj}} - p_\mathrm{max}^{3-q_\mathrm{inj}} \right) \;\; \mathrm{if}\; p \geq p_\mathrm{min}\,,\\
        \left( p_\mathrm{min}^{3-q_\mathrm{inj}} - p_\mathrm{max}^{3-q_\mathrm{inj}} \right) \;\ \mathrm{if}\; p < p_\mathrm{min}\,.
    \end{cases}
\end{aligned}
    \label{eq: steady-state solution}
\end{equation}
Thus, for low (high) momenta and Coulomb losses with $b(p) \approx p^{-2}$ ($b(p) \approx \mathrm{const.}$) the spectral slope roughly decreases by 3 (1). For bremsstrahlung losses with $b(p) \propto p$ (in the limit of high momenta) the slope stays the same and for synchrotron/IC losses with $b(p) \propto p^2$ the slope increases by 1, which is shown in Fig.~\ref{fig: CRe_steady_state}. Note that for $p_\mathrm{max} \rightarrow \infty$ the second term in brackets can be neglected and
\begin{equation*}
    f(p) \propto p^{1-q_\mathrm{inj}} / b(p) \propto p^{-q_\mathrm{inj}} \tau(p)\quad \mathrm{for}\; p \geq p_\mathrm{min}\,.
\end{equation*}
Hence, the shape of the steady-state spectrum reflects the total cooling time $\tau(p)$ and it peaks at those momenta for which $\tau(p)$ is longest. 

\subsubsection{Convergence test}

\begin{figure*}
    \centering
    \includegraphics[width=0.49\linewidth]{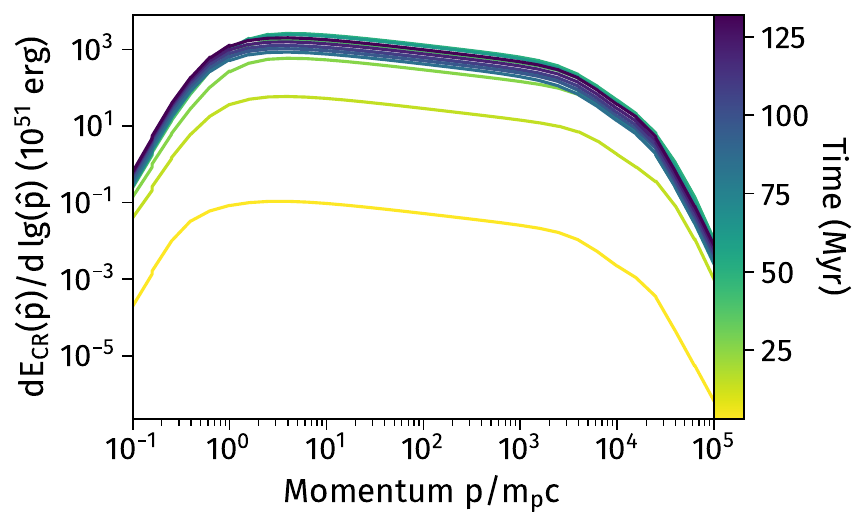}
    \includegraphics[width=0.49\linewidth]{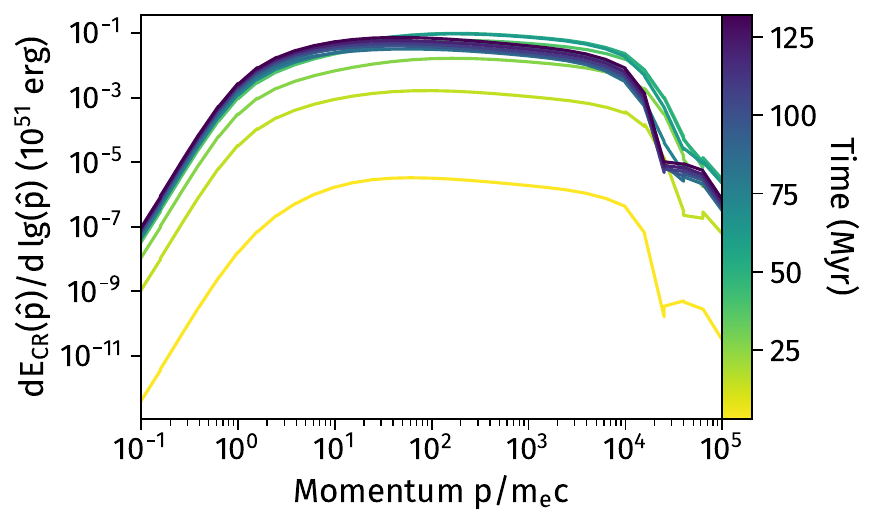}
    \caption{Injection of template CR spectra by ongoing core-collapse supernovae in a periodic ISM box. We computed the superposition of all CR spectra in the box and then plotted the logarithmic derivative of its kinetic energy (for details see App.~\ref{app: Superposition of piecewise power-law spectra}). Both the proton spectra (left) and the electron spectra (right) converge to a steady-state solution. Note that the flattening of the electron spectrum at high momenta is a feature of the models by \citet{Cristofari.etal2021}.}
    \label{fig: injection of CR spectra}
\end{figure*}

So far, we have demonstrated that for a relatively high number of bins (${\sim}10$~per dex) both the distribution function evolved with CRESCENDO and the integrated energies in each bin agree well with the corresponding analytical solution. To investigate the performance of our code more quantitatively, we reran the cooling test for CR electrons from Fig.~\ref{fig: all losses for CRe}, where we evolve the same initial spectrum for the same duration of ${\sim}50$~Myr using three different spectral resolutions, namely 1, 2, and 4 momentum bins/dex. As shown in Fig.~\ref{fig: CRe_all_losses_different_Nbins}, the final spectrum becomes noticeably discontinuous for the lowest resolution, and the relative energy error $\Delta \varepsilon_{\mathrm{rel},i}$ in each bin is of the order ${\sim}10\%$. As $N_\mathrm{bins}$ increases, the spectrum becomes smoother, and the agreement with the analytical solution is easier to see by eye, which is why we used a large number of bins in the previous tests. From Fig.~\ref{fig: CRe_all_losses_different_Nbins} one also notices that $\Delta \varepsilon_{\mathrm{rel},i}$ remains large for large momenta, because the strong spectral curvature close to the synchrotron/IC cut-off is very sensitive to the discretization error. In practice, this is only a minor problem, because these bins barely contribute to the total integrated energy of the spectrum.

Next, we reran the cooling tests for protons and electrons with all loss processes (cf. Figs.~\ref{fig: all losses for CRe} and \ref{fig: Coulomb and hadronic losses for protons}) using a large momentum range $\hat{p} \in [{5\cdot10^{-2}},\, {5\cdot10^6}]$ that we divided into different numbers of bins with resolutions ranging from $1~\mathrm{bins/dex}$ to $10~\mathrm{bins/dex}$. For the total duration $\tau_\mathrm{tot}$ of the simulation, we always use a fixed value of $\tau_\mathrm{tot} = 10^4 \cdot \Delta t_\mathrm{sub}$, where the cooling timestep $\Delta t_\mathrm{sub}$ is computed according to Eq.~\ref{eq: computation of Delta t_sub} for the highest spectral resolution, namely $10~\mathrm{bins/dex}$. In Fig.~\ref{fig: convergence test} we plot the $L_1$ norm of the relative energy and number errors (averaged by the number of bins),
\begin{equation}
    || \Delta \varepsilon||_1 = \frac{1}{N_\mathrm{bins}} \sum_{i=1}^{N_\mathrm{bins}} \Delta \varepsilon_{\mathrm{rel},i}\,, \quad || \Delta n||_1 = \frac{1}{N_\mathrm{bins}} \sum_{i=1}^{N_\mathrm{bins}} \Delta n_{\mathrm{rel},i}\,,
    \label{eq: L1 norm of relative errors}
\end{equation}
where $\Delta n_{\mathrm{rel},i}$ is computed analogously to $\Delta \varepsilon_{\mathrm{rel},i}$ from Eq.~\eqref{eq: relative energy error}. 

We ran the tests for initial power-law spectra with four different slopes, namely $q=3$, $q=4$, $q=5$ and the default value $q=4.5$. The case $q=3$ leads to an integrand ${\propto} \hat{p}^{-1}$ for the CR number~\eqref{eq: CR number integral final}, so the errors from all log-spaced bins contribute equally to $|| \Delta n ||_1$. Moreover, a case-distinction as in Eq.~\eqref{eq: case distinction of numerical integral} is required, which is numerically more challenging. The same is true for the CR energy in the non-relativistic limit ($q=5$) and the ultra-relativistic limit ($q=4$). We find that for both protons and electrons and for all initial slopes considered, the relative errors $|| \Delta \varepsilon||_1$ and $|| \Delta n||_1$ do not decrease significantly for more than ${\sim}3~\mathrm{bins/dex}$, which is still a low spectral resolution.

Additionally, Fig.~\ref{fig: convergence test} shows how the runtime $t_\mathrm{run}$ depends on the number of momentum bins (with up to $25~\mathrm{bins/dex}$) for the default case $q=4.5$, but the result is basically the same for the other slopes. For a low spectral resolution the scaling is quite different for protons and electrons due to the different energy loss processes involved. However, for a larger number of bins the scaling can be approximated by a power-law $t_\mathrm{run} \propto N_\mathrm{bins}^\alpha$ with slope $\alpha \approx2.2$. This can be derived from the obvious relation $t_\mathrm{run} \propto N_\mathrm{bins} N_\mathrm{steps}$, where the number of subcycling timesteps needed to reach the (fixed) total simulation time $\tau_\mathrm{tot}$ is $N_\mathrm{steps} = \tau_\mathrm{tot}/\Delta t_\mathrm{sub} \propto \Delta t_\mathrm{sub}^{-1}$. For our choice of parameters in the convergence test (momentum range, electron density, etc.) $\Delta t_\mathrm{sub}$ is dominated by the Coulomb cooling timescale~\eqref{eq: Delta tmax Coulomb}, because it is the shortest. One can then rewrite Eq.~\eqref{eq: Delta tmax Coulomb} as a function of $N_\mathrm{bins}$ and insert this into $t_\mathrm{run} \propto N_\mathrm{bins} \Delta t_\mathrm{sub}^{-1}(N_\mathrm{bins})$. In fact, the slope of the best-fit power-law depends weakly on $N_\mathrm{bins}$ and converges to $\alpha=2$ for a very high number of bins. For a moderate spectral resolution the runtime grows sub-quadratically for protons and super-quadratically for electrons when $N_\mathrm{bins}$ increases.

\subsection{Advanced setups with multiple particles}\label{sec: Advanced setups with multiple particles}

In these idealized setups we use initial conditions with many SPH particles and a lower spectral resolution closer to what would be used in large cosmological simulations. Here, the environmental parameters and spectral properties are not set by hand, but depend on the initial conditions and the physical processes activated during the run. Amongst those are gas cooling, star formation or magnetic field seeding, which is a first step towards full physics production runs of large astrophysical systems, like galaxies and galaxy clusters.

\subsubsection{CR seeding by supernova remnants}\label{sec: CR seeding by supernova remnats}

\begin{figure}
    \centering
    \includegraphics[width=\linewidth]{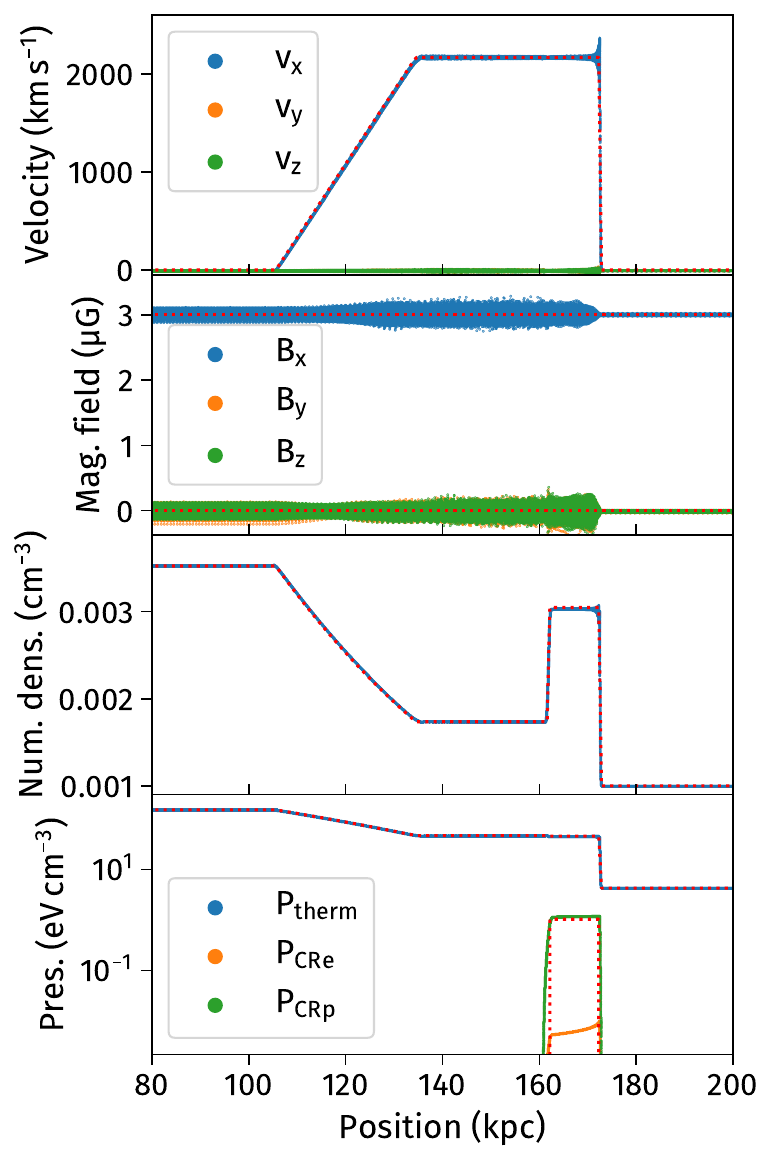}
    \caption{Sod shocktube mimicking typical ICM conditions with Mach number $\mathcal{M}=3$ after 10~Myr. CRs are injected in the post-shock region using a constant efficiency $\eta=0.1$. The red dotted line is the analytical solution by \citet{Pfrommer.etal2017}.}
    \label{fig: shocktube}
\end{figure}

To test our subgrid model for CR injection by supernova remnants in an idealized setup, we use a box with a side length of 5~kpc with periodic boundary conditions containing a cubic grid of $20^3$ gas particles with a small central over-density. The particle masses are chosen so that the average hydrogen number density is close to the star formation threshold of $1.84 \cdot 10^{-1}~\si{cm^{-3}}$. 

Shortly after the simulation is started, star formation occurs in the box centre according to the effective model by \citet{Springel.Hernquist2003} and supernova explosions inject tabulated CR spectra from \citet{Cristofari.etal2021} in the star-forming SPH particles. Thereafter, the CRs are distributed amongst neighbouring particles with an SPH-like kernel. We then compute the superposition of all spectra for every snapshot, as explained in App.~\ref{app: Superposition of piecewise power-law spectra}. The result is plotted in Fig.~\ref{fig: injection of CR spectra}, where the interplay between (non-constant) injection and cooling leads to the gradual formation of a steady-state spectrum. Note that the spectral features do not only arise from the momentum loss processes, but are also intrinsic to the template spectra, which are no simple power-laws any more.

\begin{figure*}
    \centering
    \includegraphics[width=0.49\linewidth]{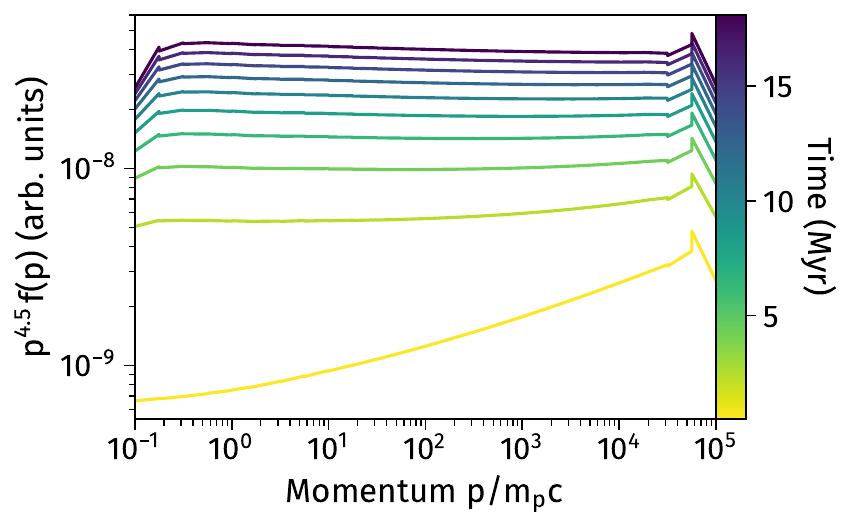}
    \includegraphics[width=0.49\linewidth]{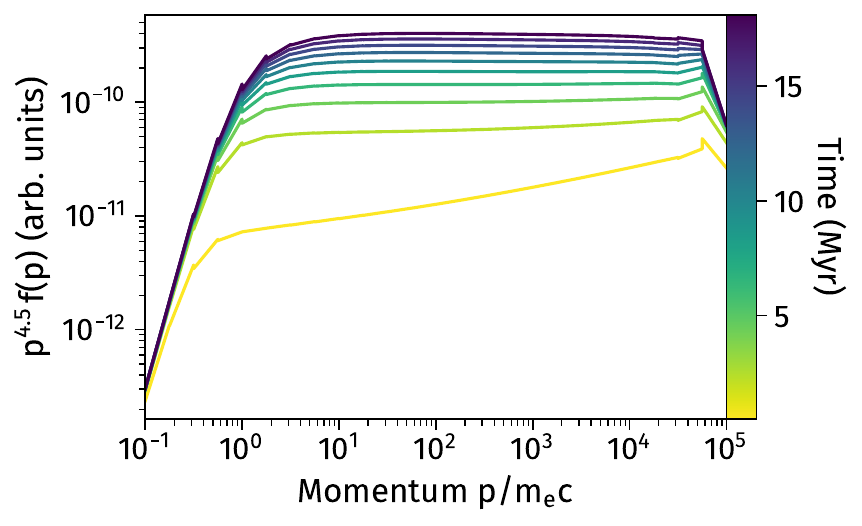}
    \caption{Superposition of all CR proton (left) and electron spectra (right) in the Sod shocktube shown in Fig.~\ref{fig: shocktube}. The spectral shape of CR protons is barely modified due to their long cooling timescale.}
    \label{fig: CR spectrum at the shock}
\end{figure*}

\subsubsection{Shocktube}

As described in our previous work \citep{Boess.etal2023, Boess.etal2024}, the shockfinder in \textsc{OpenGadget3} \citep{Beck.etal2016} detects shock waves in the simulation on the fly and injects power-law spectra for both CR protons and CR electrons at that location. The spectral slope depends on the sonic Mach number, and the normalization depends on both the Mach number and the magnetic field orientation relative to the shock front, where we use analytical fitting functions based on state-of-the-art PIC simulations. 

We set up a classical 3D Sod shocktube \citep{Sod1978} using a box with dimensions $280 \times 1 \times 1~\mathrm{kpc^3}$ and periodic boundary conditions 
containing ${\sim}6.3 \cdot 10^5$ gas particles. There is an initial jump in gas density and pressure in the middle of the tube, whereas the initial gas velocity is zero and the magnetic field is $\vb{B}=(3,0,0)^\top~\si{\mu G}$. Note that the perpendicular magnetic field vanishes, so the shock is essentially hydrodynamical. We use typical conditions of the intracluster medium (ICM) for the upstream parameters, namely a plasma number density $n = 10^{-3}~\si{cm^{-3}}$, a mean molecular weight $\mu=0.59$ (corresponding to a fully ionized H-He plasma) and a gas temperature $T=5 \cdot 10^7$~K. The downstream state is chosen so that the shock wave has a sonic Mach number $\mathcal{M} = 3$. The velocity, density and pressure of the gas after 10~Myrs as well as the CR pressure are shown in Fig.~\ref{fig: shocktube}. In order to compare this with the analytical solution from \citet{Pfrommer.etal2017} we use a constant acceleration efficiency $\eta = \varepsilon_\mathrm{CR}/\varepsilon_\mathrm{sh} = 0.1$, defined as the ratio between the total CR energy $\varepsilon_\mathrm{CR}$ injected in the post-shock region and the energy $\varepsilon_\mathrm{sh}$ dissipated at the shock. 

Since CR electrons and protons are only injected at the shock location, their pressure profiles are essentially rectangular and broaden over time. In Fig.~\ref{fig: CR spectrum at the shock} we show the superposition of all CR spectra in the shocktube at different times (similar to Fig.~\ref{fig: injection of CR spectra}). The low- and high-momentum part of the total electron spectrum are visibly affected by Coulomb and synchrotron/IC cooling, whereas the shape of the proton spectrum remains nearly unchanged, apart from some discretization noise. The reason is that the shock's evolution time (and the total simulation time) is much shorter than the cooling timescale for CR protons. In addition, CR electrons far downstream had enough time to cool radiatively, leading to a visible decline in their pressure.

\subsubsection{Constraints on timesteps}\label{sec: Constraints on timesteps}

The two setups that were investigated in the previous sections (shocktube and ISM box) also illustrate the importance of subcycling a simulation timestep. The relevant figures and additional details can be found in App.~\ref{app: Comparison of timesteps for idealized simulations}; here we summarize the main results. For the shocktube mimicking typical ICM conditions, the subcycling timestep $\Delta t_\mathrm{sub}$, which is the minimum of all cooling times according to Eq.~\eqref{eq: computation of Delta t_sub}, is always larger than the simulation timestep $\Delta t_\mathrm{sim}$ for both CR protons and CR electrons, as shown in Fig.~\ref{fig: timesteps for Sod shocktube}. Therefore, we do not have to divide $\Delta t_\mathrm{sim}$ into smaller substeps, and the total simulation runtime stays the same if we deactivate subcycling. However, for merger shocks in cosmological simulations of galaxy clusters a moderate number of electron subcycling timesteps is required, because typically $\Delta t_\mathrm{sim}$ is ${\sim}20$~times larger than in the shocktube test.

For the ISM box we have the exact opposite case, where the number of subcycling steps $N_\mathrm{sub} = \lceil \Delta t_\mathrm{sim} / \Delta t_\mathrm{sub} \rceil$ can become fairly large, especially for CR electrons (see Fig.~\ref{fig: subcycling histograms for ISM box}). Consequently, when we switch off subcycling the simulation runs ${\sim}450$ times faster, but large parts of the spectrum cannot be evolved correctly. This is shown in Fig. ~\ref{fig: electron spectra and cooling times for ISM box} together with the momentum-dependent cooling time $\tau_\mathrm{cool}(p) = | p / \dot{p}|$ for CR electrons. As expected, our CR solver fails in the momentum range where $\tau_\mathrm{cool}(p) \lesssim \Delta t_\mathrm{sim}$.

Consequently, before carrying out computationally demanding cosmological simulations with CRESCENDO one should estimate the shortest CR cooling time to be expected in the given setup and then adapt the maximum number of subcycling steps $N_\mathrm{sub,max}$ and the momentum range $[\hat{p}_\mathrm{min}, \hat{p}_\mathrm{max}]$ for which the distribution function is represented. Since the timescale for Coulomb losses at small momenta becomes very short for CR electrons, it is reasonable to only simulate CR electrons with $\hat{p} \gtrsim 10$, corresponding to a kinetic energy of ${\gtrsim} 4.6$~MeV. As mentioned in Sec.~\ref{sec: Injection of power-law spectra}, this is no severe restriction, because CR electrons with such small momenta barely contribute to the total energy budget and the non-thermal emission in wavebands of interest. We find this to be a good compromise for large-scale simulations together with the default setting $N_\mathrm{sub,max}=8$. 


\section{Conclusions}

In this paper we introduced several extensions to the on-the-fly spectral cosmic-ray (CR) solver CRESCENDO in the cosmological simulation code \textsc{OpenGadget3}. On the one hand, we implemented the seeding of physically motivated CR spectra by supernova remnants, thereby quantitatively improving the standard subgrid description, where a power-law spectrum normalized to 10\% of the canonical supernova energy $E_\mathrm{SN} = 10^{51}$~erg is injected for CR protons. The implementation is flexible in the sense that the template spectra can easily be updated in the future if better models become available, because they are stored in separate text files.

On the other hand, we implemented a number of crucial energy loss processes for both CR electrons and CR protons. Our spectral cosmic-ray solver now contains Coulomb, hadronic, bremsstrahlung, and synchrotron/IC losses. For clarity, the corresponding expressions for the loss rates, the initial momentum $p_u$ needed to compute particle and energy fluxes, and the timestep sizes were discussed in great detail in Sec.~\ref{sec: Numerical methods} and then thoroughly tested in Sec.~\ref{Sec: Numerical tests}. We illustrated the numerical robustness of our algorithm by evolving single spectra over long timescales of ${\gtrsim}10^4$ cooling times. The evolved spectra agree well with the corresponding analytical solutions, and the relative errors of CR numbers and energies already converge for a low spectral resolution of ${\sim}3~\mathrm{bins/dex}$.

As a first step towards cosmological simulations, we also simulated an ISM box and a shocktube with full CR and MHD physics. These tests are meant to demonstrate the performance of our updated CR solver in more sophisticated environments, but also show that in cosmological simulations one must make a careful compromise between the maximum number of subcycling timesteps and the momentum range for which the distribution function is numerically represented. If the cooling times become too short, which is a problem especially for CR electrons, then the CR solver might fail, or the number of subcycling steps might become prohibitively large.

Last but not least, we discuss a few interesting numerical subtleties in the appendices, namely 1) having a flexible cut-off momentum prevents an artificial spectral curvature; 2) the slope reconstruction might fail if analytical approximations are used for the integration boundaries of the flux integrals; and 3) hadronic losses might be interpreted as a continuous or catastrophic process.

\begin{acknowledgements}
The authors would like to thank the anonymous referee for the constructive feedback and insightful comments, which significantly improved the quality of the manuscript. DK wants to thank Giovanni Morlino, Pierre Cristofari and Iurii Sushch for very detailed and insightful discussions on CR acceleration via supernova remnants and for providing their cosmic-ray spectra. DK also thanks Damiano Caprioli and Philipp Girichidis for interesting discussions on diffusive shock acceleration and numerical methods. 
DK, IK and KD acknowledge support by the COMPLEX project from the European Research Council (ERC) under the European Union’s Horizon 2020 research and innovation program grant agreement ERC-2019-AdG 882679.
KD was supported by the Excellence Cluster ORIGINS which is funded by the Deutsche Forschungsgemeinschaft (DFG, German Research Foundation) under Germany's Excellence Strategy -- EXC-2094 -- 390783311.
LMB is supported by NASA through grant 80NSSC24K0173 and NSF through grant AST-2510951.
IK was supported by the Simons Foundation via the Simons Investigator Award to A. A. Schekochihin.\\

Used software: Matplotlib \citep{Hunter2007}, numpy \citep{Harris.etal2020}, Scipy \citep{Virtanen.etal2020}, g3read (\url{https://github.com/aragagnin/g3read}), GNU Scientific Library \citep{Galassi.etal2021}
\end{acknowledgements}

%

\bibliographystyle{aa}
\bibliography{Literature}

\begin{appendix}
\nolinenumbers




\section{Numerical computation of improved energy and pressure integrals} \label{app: Numerical computation of improved energy and pressure integrals}

In principle, the integrals for CR energy and pressure (cf. Eqs.~\eqref{eq: CR energy integral final} and \eqref{eq: CR pressure integral final}) could be written in terms of the hypergeometrical function (or the incomplete beta function). However, it is more convenient to approximate these integrands as piecewise power-laws and then integrate them analytically \citep[e.g.][]{BaldacchinoJordan.etal2025}. More precisely, in a given interval $[x_i, x_{i+1}]$ we approximate an arbitrary function $g(x)$ as
\begin{align}
    g(x) = c_i x^{-q_i}\,, \quad x \in[x_i, x_{i+1}]\,.
    \label{eq: piecewise power-law representation}
\end{align}
The slope $q_i$ is estimated from a linear interpolation in log-log space,
\begin{align*}
    q_i = -\frac{\lg(g_{i+1}) - \lg(g_i)}{\lg(x_{i+1}) - \lg(x_i)} \,
\end{align*}
which is of course exact for power-laws. Inverting equation~\eqref{eq: piecewise power-law representation} then gives the normalization constant $c_i$,
\begin{align*}
    c_i = \left. g(x) x^{q_i} \right|_{x=x_\mathrm{mid}}\,,
\end{align*}
where $x_\mathrm{mid} = \sqrt{x_i \,x_{i+1}}$ is the geometric mean of the bin boundaries (equivalent to the bin centre in log space). Evaluating the r.h.s at the ``bin middle'' $x_\mathrm{mid}$ instead of the left boundary $x_i$ provides a better approximation to the analytical function. 

The analytical solution requires a case distinction:
\begin{align}
    I =\int \limits_{x_1}^{x_2} x^{-q} \dd{x} = \begin{cases}
       \displaystyle I_1 \coloneqq \frac{x_2^{1-q} - x_1^{1-q}}{1-q} \quad &\mathrm{for}\; q \neq 1\,, \\
       \displaystyle I_2 \coloneqq \ln(x_2/x_1) \quad &\mathrm{for}\; q = 1\,.
    \end{cases}
    \label{eq: case distinction of numerical integral}
\end{align}
Numerically, we interpolate between the two functions $I_1$ and $I_2$ if $\abs{q -1 } < \varepsilon_\mathrm{soft}$ (otherwise we just use $I_1$), where $\varepsilon_\mathrm{soft}$ is the slope softening with default value $\varepsilon_\mathrm{soft}=10^{-6}$:
\begin{equation*}
    I \approx I_1 \frac{\abs{q-1}}{\varepsilon_\mathrm{soft}} + I_2 \left( 1 - \frac{\abs{q-1}}{\varepsilon_\mathrm{soft}}\right) \quad \mathrm{for}\; \abs{q - 1} < \varepsilon_\mathrm{soft} \,.
\end{equation*}
Conversely, $\abs{1 - q}$ can be quite large, because $q \in [-20,20]$ by default in all simulation setups. Hence, to avoid overflow errors, $I_1$ and $I_2$ are always computed with double precision.

\begin{figure*}
    \centering
    \includegraphics[width=\textwidth]{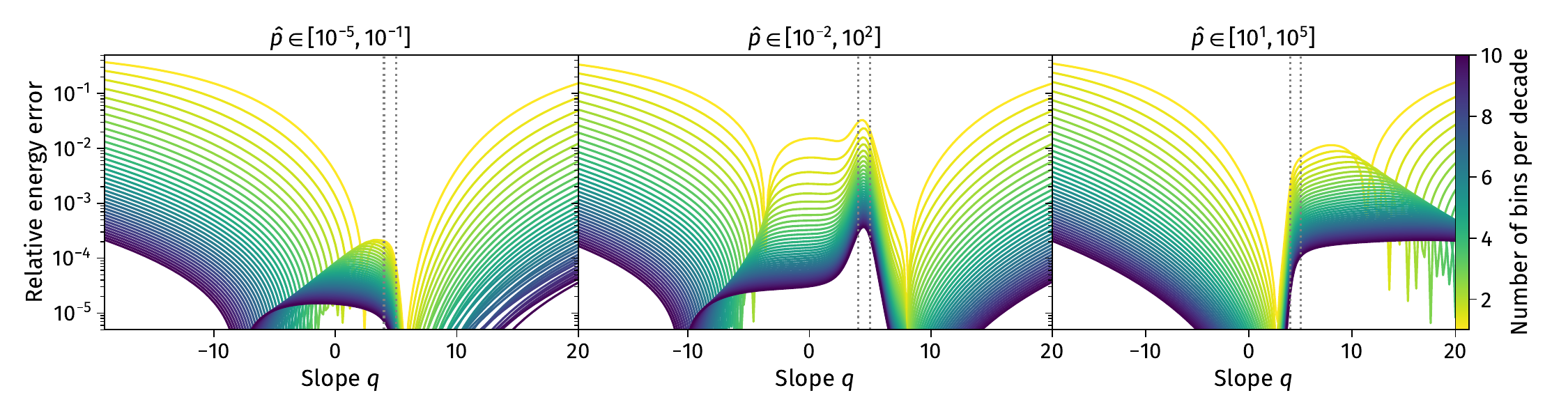}
    \caption{Relative error of the energy integral~\eqref{eq: CR energy integral final} for various slopes, where the three integration intervals are divided into different numbers of bins. The gray dotted lines show the physically important slope range $q \in [4,5]$.}
    \label{fig: energy_integrals_comparison}
\end{figure*}

To test the accuracy of this power-law approximation we compute the integrated CR energy~\eqref{eq: CR energy integral final} for various slopes $q \in [-20,20]$ in the three momentum ranges $\hat{p} \in [10^{-5}, 10^{-1}]$, $\hat{p} \in [10^{-2}, 10^2]$ and $\hat{p} \in [10^{1}, 10^5]$, corresponding to the non-relativistic and the ultra-relativistic regime and the transition in between. In Fig.~\ref{fig: energy_integrals_comparison} we plot the relative error with respect to the exact value of the integral for every slope and for different numbers of bins. It is evident that the relative error decreases with increasing number of bins and that it tends to be larger for certain slopes. Nevertheless, already for a moderate discretization of ${\sim} 3~\mathrm{bins/dex}$ the relative error is smaller than one percent for most of the practically relevant slopes. We found that the same conclusion also holds for the pressure integral~\eqref{eq: CR pressure integral final}.

Finally, we note that the exact value of the integral, which we used for calculating the relative error, can be computed numerically to arbitrary precision by increasing the number of grid points on which the integrand is evaluated. However, it is much faster to perform a logarithmic variable substitution and to integrate the transformed function. More precisely, if $g(p)$ is an arbitrary function, then the substitution $p \rightarrow s = \ln(p)$ leads to:
\begin{align*}
	\int \limits_{p_1}^{p_2} g(p) \dd{p} = \int \limits_{\ln(p_1)}^{\ln(p_2)} g(e^s) e^s \dd{s} \approx \frac{1}{2} \ln(\frac{p_2}{p_1}) \, \left(g(p_1)p_1 + g(p_2)p_2 \right)\,,
\end{align*}
where in the last step the trapezoidal rule was used. For functions that behave like piecewise power-laws this substitution works very well.

\section{Summary of two-moment approach}\label{app: Summary of two-moment approach}

The formalism that we use to evolve the distribution function of protons and electrons is outlined in our previous work \citet{Boess.etal2023} and in many other papers \citep[e.g.][]{Miniati2001, Girichidis.etal2020, Ogrodnik.etal2021, Hanasz.etal2021, Hopkins.etal2022a, BaldacchinoJordan.etal2025}. For completeness, we summarize the main steps in this appendix. The starting point is the standard evolution equation for a nearly isotropic CR distribution function $f(\vb{x}, p, t)$ on macroscopic scales in the strong-scattering regime \citep{Skilling1975, Schlickeiser2002, Zank2014}:
\begin{equation}
\begin{aligned}
	\pdv{f(\mathbf{x}, p, t)}{t} &= - \vb{u} \vdot \grad{f}(\mathbf{x}, p, t) + \div( \tensorfont{D}(\vb{x}, p, t) \grad{f(\mathbf{x}, p, t)})\\
    &+ \left(\frac{1}{3} \div{\mathbf{u}}\right) p \frac{\partial f(\mathbf{x}, p, t)}{\partial p} \\
	&+\frac{1}{p^2} \frac{\partial}{\partial p}\left(p^2 \sum_j b_j(p) f(\mathbf{x},p,t) \right) \\ 
    &+ j(\mathbf{x}, p, t)  \,. 
\end{aligned}
    \label{eq: CR transport equation}
\end{equation}
The terms on the r.h.s describe 1) advection/streaming, where $\vb{u}(\vb{x},t) = \vb{v}(\vb{x},t) + \vb{v}_\mathrm{str}(\vb{x},t)$ is the sum of the mean gas velocity $\vb{v}$ (bulk motion) and the streaming velocity $\vb{v}_\mathrm{str}$ \citep{Ruszkowski.Pfrommer2023}, 2) spatial diffusion with a diffusion tensor $\tensorfont{D}(\vb{x},p,t)$, 3) adiabatic compression/expansion, 4) continuous losses represented by the individual loss rates $-\dot{p}_\mathrm{tot} = b_\mathrm{tot} = \sum_j b_j$ and 5) CR injection via a source term $j(\vb{x},p,t)$. In this paper, we focus especially on the continuous loss processes and CR injection by supernova remnants; details on the spatial transport will be covered in future work. We emphasize that the distribution function in Eq.~\eqref{eq: CR transport equation} is expressed in mixed reference frames. The momentum $p$ is measured in the rest frame of the gas, whereas the spatial coordinates $\vb{x}$ are measured in the observer's rest frame. The relative velocity between these two reference frames is the mean gas velocity $\vb{v}$. We also note that \citet{Miniati2001} expressed Eq.~\eqref{eq: CR transport equation} in terms of the comoving coordinates $\vb{x}_\mathrm{com} = \vb{x} / a(t)$, where $a(t)$ is the cosmological scale factor. Although this formulation is most appropriate for a cosmological simulation code, we convert all ambient quantities (e.g. gas density or magnetic field) into physical units before inserting them into Eq.~\eqref{eq: CR transport equation}.

Due to the Lagrangian nature of our code we add the advection term to the l.h.s of Eq.~\eqref{eq: CR transport equation} and express it in terms of the advective time derivate $\mathrm{d}/\mathrm{d}t = \partial/\partial t + \vb{u} \vdot \grad$. Then we multiply both sides with $4 \pi p^2/ \rho(\vb{x},t )$ and integrate from $p_i$ to $p_{i+1}$ to obtain an evolution equation for the CR number per unit mass in the $i$-th bin:
\begin{align}
	\dv{n_i}{t} 
	&= \frac{1}{\rho} \div( \langle \tensorfont{D} \rangle_{n,i} \grad (\rho n_i))  + \left[ -\left( \dv{p}{t} \right)_\mathrm{tot} \frac{4 \pi}{\rho} p^2 f \right]^{p_{i+1}}_{p_i}  + Q_i\,,
    \label{eq: dn/dt full}
\end{align}
where we dropped the function arguments for brevity and summarized all continuous loss processes in $\dot{p}_\mathrm{tot}$ (for an overview see App.~\ref{app: Summary of continuous loss processes}). The number-averaged diffusion tensor $\langle \tensorfont{D} \rangle_{n,i}$ has components
\begin{align*}
	 \langle D_{jk}(\vb{x},t) \rangle_{n,i} &\coloneqq \frac{\int_{p_i}^{p_{i+1}} p^2 D_{jk}(\vb{x}, p, t) \, \nabla_k f(\vb{x},p, t) \dd{p}}{\int_{p_i}^{p_{i+1}} p^2 \, \nabla_k f(\vb{x},p ,t) \dd{p}}\,, \\ 
\end{align*}
and the corresponding source term is
\begin{align*}
	 Q_i(\vb{x},t) &\coloneqq \frac{1}{\rho} \int \limits_{p_i}^{p_{i+1}} 4 \pi p^2 j(\vb{x}, p, t) \dd{p}\,.
\end{align*}

On the other hand, multiplying \eqref{eq: CR transport equation} with $4 \pi p^2 T(p)/ \rho(\vb{x},t)$ and integrating from $p_i$ to $p_{i+1}$ gives the evolution of the kinetic energy per mass in the $i$-th bin:
\begin{equation}
\begin{aligned}
	\dv{\varepsilon_i(\mathbf{x}, p, t)}{t} 
	&= \frac{1}{\rho} \div( \langle \tensorfont{D} \rangle_{\varepsilon,i} \grad( \rho \varepsilon_i))	+ S_i - R \varepsilon_i \\
    &+ \left[ -\left( \dv{p}{t} \right)_\mathrm{tot} \frac{4 \pi}{\rho} p^2 T(p) f \right]^{p_{i+1}}_{p_i} \,. \\
    \label{eq: de/dt full}
\end{aligned}
\end{equation}
The energy-weighted diffusion coefficient has components and source term are
\begin{align*}
	\langle D_{jk}(\vb{x},t) \rangle_{\varepsilon,i}  &= \frac{\int_{p_i}^{p_{i+1}} p^2 T(p) \, D_{jk}(\vb{x},p,t)\, \nabla_k f(\vb{x},p, t)   \dd{p}}{\int_{p_i}^{p_{i+1}}  p^{2} T(p) \, \nabla_k f(\vb{x},p, t)\dd{p}}\,, \\
\end{align*}
and the corresponding source term is
\begin{align*}
	S_i(\vb{x},t) &= \frac{1}{\rho} \int \limits_{p_i}^{p_{i+1}} 4 \pi p^2 T(p) j(\vb{x}, p, t)\, \dd{p} \,.
\end{align*}
An important difference with respect to Eq.~\eqref{eq: dn/dt full} for the evolution of $n_i$ is the emergence of a new term,
\begin{align*}
	R_i(\vb{x},t) &= \frac{1}{\varepsilon_i \rho} \int \limits_{p_i}^{p_{i+1}} b_\mathrm{tot}(p) \, 4 \pi p^2 f \dv{T(p)}{p} \dd{p} \\
    &= - \frac{1}{\varepsilon_i \rho} \int \limits_{p_i}^{p_{i+1}} 4 \pi p^2 f \dot{T}(p) \dd{p} \,,
\end{align*}
which can be interpreted as relative energy change (per density). After inserting Eq.~\eqref{eq: piecewise power-law representation of f} for $f(p)$ this can be written purely in terms of the integrals~\eqref{eq: CR energy integral final} and \eqref{eq: CR pressure integral final}, where the prefactors cancel out:
\begin{align*}
    R(\vb{x},t) &= -\frac{1}{\varepsilon_i \rho} \int \limits_{p_i}^{p_{i+1}} 4 \pi p^2 f \dot{p}_\mathrm{tot} \frac{pc^2}{\sqrt{(pc)^2 + (mc^2)^2}} \dd{p} \\
    &= - \frac{4 \pi (mc)^4 c f_i \hat{p}_i^{q_i}}{\varepsilon_i \rho} \int \limits_{\hat{p}_i}^{\hat{p}_{i+1}}  \dot{\hat{p}}_\mathrm{tot} \frac{\hat{p}^{3-q_i}}{\sqrt{\hat{p}^2 + 1}} \dd{\hat{p}} \\
    &= \left( \int \limits_{\hat{p}_i}^{\hat{p}_{i+1}} \left(\sqrt{\hat{p}^2+1} -1 \right) \hat{p}^{2-q_i} \dd{\hat{p}} \right)^{-1} \int \limits_{\hat{p}_i}^{\hat{p}_{i+1}} b_\mathrm{tot}(\hat{p}) \frac{\hat{p}^{4-(q_i+1)}}{\sqrt{\hat{p}^2 + 1}} \dd{\hat{p}} .
\end{align*}

\section{Importance of updating the spectral cut-off} \label{app: Importance of updating the spectral cut-off}

\begin{figure}
    \centering
    \includegraphics[width=\linewidth]{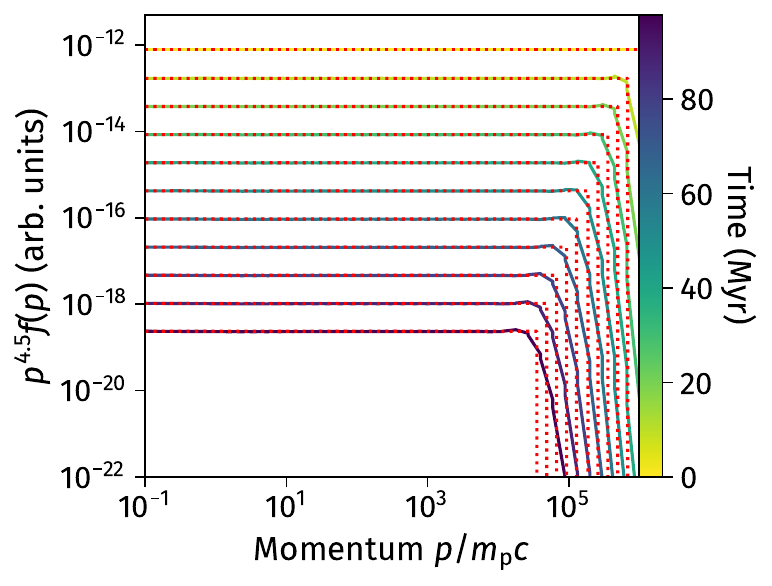}
    \caption{Same as Fig.~\ref{fig: CRp adiabatic expansion test}, but the spectral cut-off is not updated, which leads to an artificial spectral curvature at high momenta.}
    \label{fig: CRp adiabatic expansion test without updating cut-off}
\end{figure}

We show the importance of updating the spectral cut-off in every timestep and therefore having an upper momentum boundary that is not fixed. For this we adiabatically cool a CR proton spectrum, where the new cut-off after a timestep $\Delta t$ is given by:
\begin{align*}
 \hat{p}_\mathrm{cut,new} &= \hat{p}_\mathrm{cut,old} \, \exp(-B_\mathrm{ad} \Delta t) = \hat{p}_\mathrm{cut,old} \, \exp\left(- \frac{1}{3} (\div{\vb{u}})\, \Delta t\right)\,.
\end{align*}
From the analytical solution Eq.~\ref{eq: analytical solution for adiabatic losses for power-law} it is clear that the slope of the initial power-law spectrum should be conserved. Numerically, this is only true if the cut-off is updated as in Fig.~\ref{fig: CRp adiabatic expansion test}; otherwise an artificial spectral curvature emerges for large momenta, where the bins have different slopes, as shown in Fig.~\ref{fig: CRp adiabatic expansion test without updating cut-off}. The same is true for hadronic and synchrotron/IC losses (in the case of electrons). Note that the fixed lower cut-off is less problematic, because we allow an influx of particles whenever CRs gain energy (e.g. via an adiabatic compression) mimicking the acceleration of particles from the thermal pool. In contrast, an influx of particles at the upper momentum boundary is entirely unphysical.

\section{Computation of the initial momentum for CR fluxes} \label{app: Computation of the initial momentum for CR fluxes}

\begin{figure*}
    \centering
    \includegraphics[width=0.49 \textwidth]{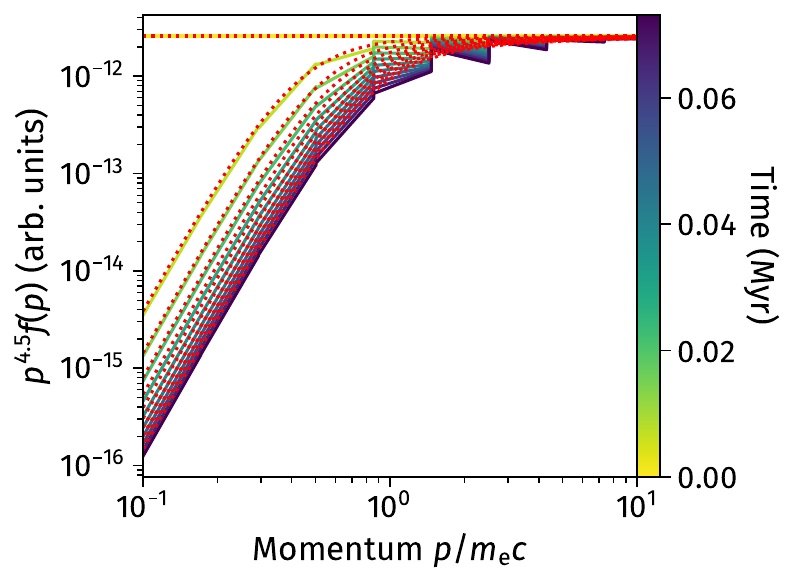}
    \includegraphics[width=0.49 \textwidth]{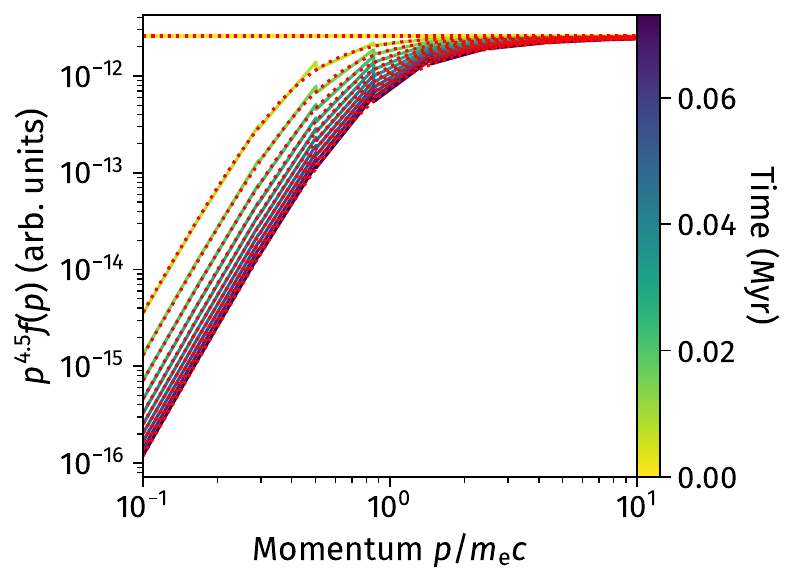}
    \caption{Freely cooling electron spectrum affected only by Coulomb losses. In the left-hand panel the initial momentum $p_u$ (needed for the computation of number and energy fluxes) was calculated by using the approximation~\eqref{eq: pu Coulomb Winner}, whereas in the right-hand panel the implicit equation~\eqref{eq: pu Coulomb from implicit equation} was solved with Brent's method. Only the latter method is fully consistent with the loss rate~\eqref{eq: approximation for dp/dt Coulomb}, so the slope reconstruction and the agreement with the analytical solution (red dotted line) is much better.}
    \label{fig: Coulomb losses for CRe with pu computed in different ways}
\end{figure*}

In Fig.~\ref{fig: Coulomb losses for CRe with pu computed in different ways} we show a freely cooling electron spectrum with only Coulomb losses switched on. As discussed in Sec.~\ref{sec: Numerical methods} the initial momentum $p_u$ is an integration boundary needed to compute the CR number and energy fluxes $\Phi_{n,i}$ and $\Phi_{\varepsilon,i}$ (see Eqs.~\eqref{eq: CR number flux} and \eqref{eq: CR energy flux}). If we use the analytical approximation~\eqref{eq: pu Coulomb Winner} for $p_u$, then the reconstructed slopes in the low-momentum regime are unphysical and strongly deviate from the analytical solution (left panel). The problem is that Eq.~\eqref{eq: pu Coulomb Winner} is no longer fully consistent with the assumed loss rate~\eqref{eq: approximation for dp/dt Coulomb}, which is needed to compute the relative energy change $R_i$ according to Eq.~\eqref{eq: relative energy change term R}. Hence, the updated energy $\varepsilon_{\mathrm{CR},i}$~\eqref{eq: updated e} is systematically wrong, because incompatible values for $\Phi_{\varepsilon,i}$ and $R_i$ are used. Due to this inconsistency the reconstructed slopes $q_i$ (cf. Eq.~\eqref{eq: update of spectral slope}) deviate from their true values already after a few cooling times, as shown in the left panel of Fig.~\ref{fig: Coulomb losses for CRe with pu computed in different ways}. In contrast, solving the implicit equation~\eqref{eq: pu Coulomb from implicit equation} for $p_u$ with a root finder allows a more accurate update of $\varepsilon_{\mathrm{CR},i}$ and therefore a better slope reconstruction. The same is true for all other loss processes (for hadronic losses see the right panel in Fig~\ref{fig: simulation of 3 different approaches for hadronic losses}), where an analytical expression for $p_u$ is used, that is no exact solution of \eqref{eq: IVP for b(p)}. Consequently, a stable slope reconstruction is worth the higher cost of an iterative method for computing $p_u$.

\section{Continuous and catastrophic momentum loss processes}\label{app: Continuous and catastrophic loss processes}

The general differential equation for continuous momentum loss processes $\dot{p} = -b(p)$ can be derived by writing the continuity equation in momentum space in spherical coordinates, where only the term containing $\dot{p}$ remains, because all loss processes considered here only change the magnitude, but not the direction of the momentum, i.e. $\dot{\theta}=\dot{\varphi}=0$:
\begin{align*}
    0 &= \pdv{f(\vb{x}, p,t)}{t} + \grad_{\vb{p}} \vdot (\dot{\vb{p}}f(\vb{x}, p,t)) \\
    &= \pdv{f(\vb{x}, p,t)}{t} + \frac{1}{p^2} \pdv{p} (p^2 \dot{p} f(\vb{x}, p,t))\\
    &= \pdv{f(\vb{x}, p,t)}{t} - \frac{1}{p^2} \pdv{p} (p^2 b(p) f(\vb{x}, p,t)) \,.
\end{align*}
Note that by introducing the ``one-dimensional distribution function'' $f_\mathrm{1D} \coloneqq 4 \pi p^2 f$ this can be written as a 1D advection equation in momentum space \citep[e.g.][]{Hopkins.etal2022a}:
\begin{align}
   \pdv{f_\mathrm{1D}(\vb{x}, p, t)}{t} = \pdv{p} ( b(p) f_\mathrm{1D}(\vb{x},p,t)) \,.
       \label{eq: momentum space advection for f1D}
\end{align}
Particle number conservation is manifest, since we started from a continuity equation, but it can also be seen from this equivalent advection-like structure or by directly integrating both sides of Eq.~\eqref{eq: momentum space advection for f1D} over the entire phase space: 
\begin{align*}
    \dot{N} &= \pdv{t} \int_{\mathbb{R}^3} \dd{\vb{x}} \int \limits_0^\infty \dd{p} f_\mathrm{1D}(\vb{x}, p,t) \\
    &= \int_{\mathbb{R}^3} \dd{\vb{x}} \int \limits_0^\infty \dd{p} \pdv{p} ( b(p) f_\mathrm{1D}(\vb{x}, p,t)) \\
    &= \int_{\mathbb{R}^3} \dd{\vb{x}}  b(p) f_\mathrm{1D}(\vb{x}, p,t) \big|_0^\infty = 0 \,.
\end{align*}
The boundary terms vanish, because $b(p=0)=0$ (no momentum can be lost anymore) and $\lim_{p\rightarrow \infty} f_\mathrm{1D}(\vb{x}, p, t)=0$ (otherwise the integral over $p$ would not be finite).
A general solution is given by
\begin{align*}
    f_\mathrm{1D}(\vb{x}, p,t) = \frac{1}{b(p)}\, g\left(\vb{x}, t + \int_{p_0}^p \frac{\dd{s}}{b(s)} \right) \,,
\end{align*}
where $g$ is an arbitrary function satisfying the initial condition $f(\vb{x}, p_0, t) = g(\vb{x}, t)/b(p_0)$.
The solution of~\eqref{eq: momentum space advection for f1D} with the more useful initial condition $f_\mathrm{1D}(\vb{x}, p,t_0) = f_\mathrm{ini}(\vb{x}, p)$ directly follows from particle number conservation for a freely cooling spectrum \citep{Sarazin1999, Winner.etal2019}\,,
\begin{align*}
    f_\mathrm{1D}(\vb{x}, p,t) = f_\mathrm{ini,1D}(\vb{x}, p_u) \frac{b(p_u)}{b(p)} \,, \quad \int \limits_p^{p_u} \frac{\dd{s}}{b(s)} = t-t_0\,,
\end{align*}
where the second equation implicitly defines the initial momentum $p_u$ as before in Eq.~\eqref{eq: implicit definition of pu}. By taking the momentum and time-dependence of $p_u$ into account, this straightforwardly translates into the analytical solution~\eqref{eq: analytical solution for f due to loss processes} for the three-dimensional distribution function,
\begin{equation*}
    f(\vb{x}, p,t) = f_\mathrm{ini}(\vb{x}, p_u(p, t)) \frac{p_u^2(p, t)}{p^2} \frac{b(p_u(p,t))}{b(p)}\,.
\end{equation*}

In contrast, for catastrophic losses the rate at which particles are destroyed is the product of their number and the collision rate $\nu_\mathrm{coll}= v n_\mathrm{target} \sigma $, which depends on the CR velocity $v$, the number density of target particles $n_\mathrm{target}$ and the cross section $\sigma$. In terms of the distribution function this reads
\begin{align*}
    -\pdv{f(\vb{x},p,t)}{t} &= \nu_\mathrm{coll}(\vb{x},p,t)\, f(\vb{x},p,t) \\
    &= v(p) \sigma(p) n_\mathrm{target} (\vb{x} ,t) \, f(\vb{x},p,t) \,.
\end{align*}
By ignoring the time-dependence of the density of target particles in the background plasma this can be written in the instructive form (that could also be used for radioactive CR nuclei with decay time $\tau_\mathrm{cat}$)
\begin{equation}
        \pdv{f(\vb{x},p,t)}{t} =- \frac{f(\vb{x},p,t)}{\tau_\mathrm{cat}(\vb{x},p)}\,, 
        \;\; \tau_\mathrm{cat}^{-1}(\vb{x},p) \equiv v(p) \sigma(p) n_\mathrm{target}(\vb{x}) \,.
\label{eq: general form of df/dt for catastrophic losses}
\end{equation}
For a given initial condition $f_\mathrm{ini}(\vb{x}, p) = f(\vb{x},p,t_0)$ this results in the familiar exponentially decaying solution:
\begin{align*}
        f(\vb{x},p,t) = f_\mathrm{ini}(\vb{x},p) \exp\left( - \frac{t-t_0}{\tau_\mathrm{cat}(\vb{x},p)} \right)\,. 
\end{align*}

\section{Additional details about continuous momentum loss processes} \label{app: Summary of continuous loss processes}

\subsection{Overview}

Before going into some technical details, we briefly summarize all momentum loss processes discussed in Sec.~\ref{sec: Energy loss processes}. For electrons we consider adiabatic, Coulomb, bremsstrahlung, synchrotron and inverse Compton (IC) losses:
\begin{align*}
	{b}_{\mathrm{tot,e}}(\hat{p})
    = B_\mathrm{ad}\, \hat{p} + B_\mathrm{Coul,e}(\hat{p}^{-2} +1) + B_\mathrm{brems}(\hat{p}^{-1} + \hat{p}) + B_\mathrm{syn+IC}\,{\hat{p}}^{2}.
\end{align*}
On the other hand, for protons we consider adiabatic, Coulomb and hadronic losses:
\begin{align*}
	 {b}_{\mathrm{tot,p}}(\hat{p})
    &= B_\mathrm{ad}\, \hat{p} + B_\mathrm{Coul,p}(\hat{p}^{-2} +1)\\
    &\quad + B_\mathrm{had}\left(\sqrt{\hat{p}^2 +1} -1 \right)H(\hat{p} - \hat{p}_\mathrm{thr})\,.
\end{align*}
The prefactors are:
\begin{equation}
    \begin{aligned}
    B_\mathrm{ad} &=  \frac{1}{3} (\div{\vb{u}})\,, \\
    B_\mathrm{Coul,e} &= \frac{\omega_{\mathrm {pl}}^2 e^2}{ m_\mathrm{e} c^3 }\,  \ln \left(\frac{m_\mathrm{e}c^2}{\hbar \omega _\mathrm{pl}}\right)\,,  \\ 
     B_\mathrm{Coul,p} &= 0.9\, \frac{\omega_{\mathrm {pl}}^2 e^2}{ m_\mathrm{p} c^3 }\, \ln \left(\frac{2 m_\mathrm{e} c^2}{\hbar \omega_{\mathrm {pl}} }\right)\,, \\
    B_\mathrm{brems} &= \frac{3 \alpha _\mathrm{ fs} \sigma _\mathrm{ T} c \bar{g}}{2 \pi} n_\mathrm{e} \frac{4 -Y}{2-Y}\,,\\
    B_\mathrm{syn+IC} &= \frac{4}{3} \frac{{\sigma }_\mathrm{T}}{m_\mathrm{e}c}({u}_{B}+{u}_{\mathrm{rad}})\,, \\
    B_\mathrm{had} &= K_\mathrm{p} c n_\mathrm{N} \sigma_\mathrm{pp} \,.
    \end{aligned}
    \label{eq: summary of B-parameters}
\end{equation}
Note that the radiative loss processes for protons are suppressed by a factor $(m_\mathrm{p}/m_\mathrm{e})^2$, and therefore neglected, because the prefactors $B_\mathrm{brems}$ and $B_\mathrm{syn+IC}$ all depend on the Thomson cross section $\sigma_\mathrm{T} \propto 1/m^2$.

\begin{figure*}
    \centering
    \includegraphics[width=0.49\linewidth]{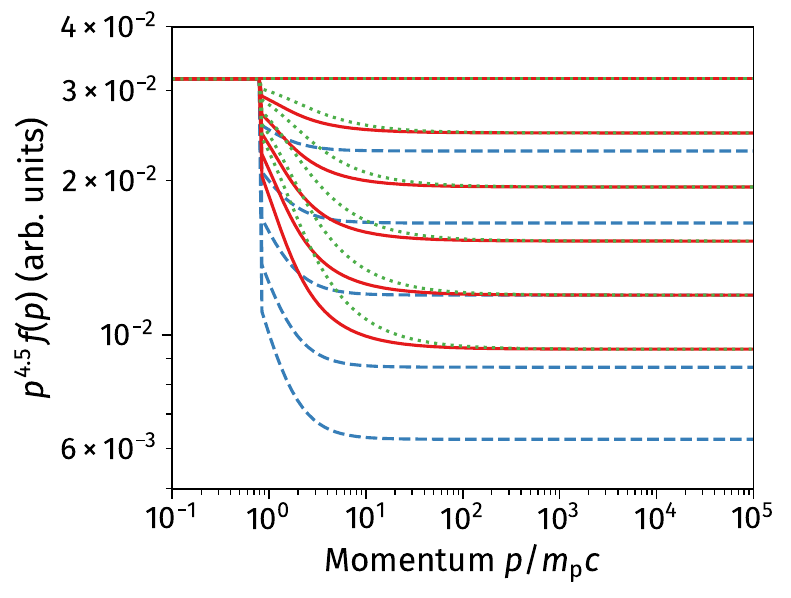}
    \includegraphics[width=0.49\linewidth]{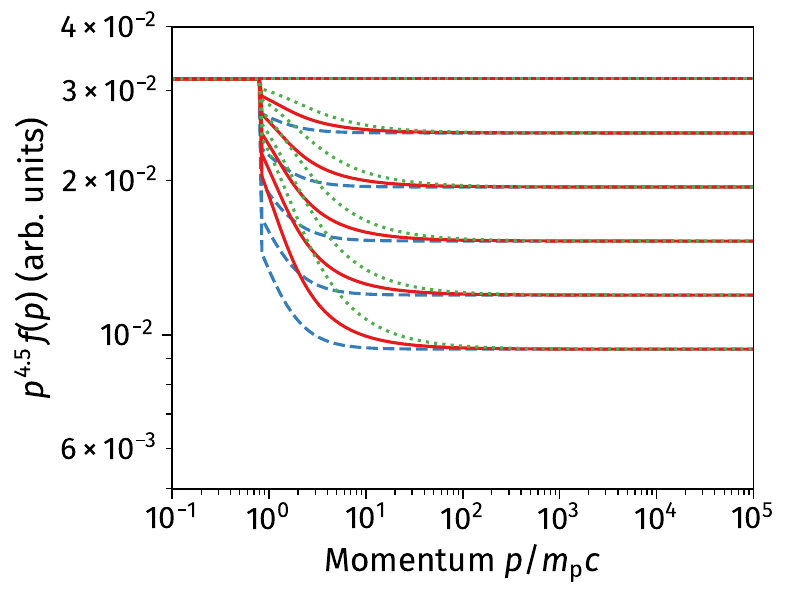}
    \caption{A freely cooling CRp spectrum with initial condition $f_\mathrm{ini}(p) = f_0(p/p_\mathrm{min})^{-4.5}$ undergoes hadronic losses for 500~Myr, where the spectrum is plotted every 100~Myr (time increases from top to bottom). We use three different numerical approaches to evolve $f_\mathrm{ini}(p)$ and plot the semi-analytical solutions (using a standard ODE solver). The blue dashed line is the catastrophic approach~\eqref{eq: analytical solution for catastrophic hadronic losses}. For the continuous approach we use Eq.~\eqref{eq: analytical solution for f due to loss processes} to compute $f(p,t)$, where we use the implicit equation~\eqref{eq: pu hadronic implicit} for $p_u$ (green dotted line) and the numerical approximation~\eqref{eq: analytical approximation for pu_hadronic} (red solid line). Left: The same cross section $\sigma_\mathrm{pp}=30$~mb is used for all lines. Right: the cross section has been rescaled according to Eq.~\eqref{eq: rescaled cross section for hadronic losses} for the catastrophic approach (blue dashed lines).}
    \label{fig: comparison between continouos and catastrophic hadronic losses}
\end{figure*}
\subsubsection{Coulomb losses}

A fast, charged particle (here either a CR proton or a CR electron) in a fully ionized plasma interacts with the free plasma electrons through Coulomb scatterings. The net effect is that the particle gradually loses some of its momentum, which is effectively transferred to the surrounding plasma. \citet{Gould1972b} provided a detailed derivation of the stopping power for a CR proton under the assumption that its recoil can be neglected, so it is only applicable for $\gamma_\mathrm{CRp} \lesssim m_\mathrm{p}/m_\mathrm{e}$, which corresponds to a critical proton energy of $1.7$~TeV. This simplification does not impose severe restrictions, because at such high energies Coulomb losses are no longer dominant anyway. The momentum loss rate is given by:
\begin{equation}
\begin{aligned}
    \left( \dv{p}{t} \right)_\mathrm{Coulomb} &= - \frac{\omega_{\mathrm {pl}}^2 e^2}{ c^2 \beta^2}\left[ \ln \left(\frac{2 m_\mathrm{e} c p \beta}{\hbar \omega_{\mathrm {pl}} m_\mathrm{p} }\right) - \frac{\beta ^2}{2} \right], \label{eq: bCoulomb according to Gould}
		\\
        \omega _\mathrm{pl} &= \sqrt{4 \pi e^2 n_\mathrm{e}/m_\mathrm{e}\, }\,, \quad \beta(p) = \frac{p/m_\mathrm{p}c}{ \sqrt{(p/m_\mathrm{p}c)^2 + 1}}\,, 
    \end{aligned}
\end{equation}
    where $\omega_\mathrm{pl}$ is the plasma frequency depending on the number density $n_\mathrm{e}$ of free electrons. The dimensionless proton velocity $\beta=v_\mathrm{CRp}/c$ should be understood as a function of the proton's momentum $p$ and since Coulomb losses are most relevant for low energies, we do not use the approximation $\beta \approx 1$. The prefactor is also often written in terms of the Thomson cross section $\sigma_\mathrm{T}$:
\begin{align*}
     \frac{\omega_{\mathrm {pl}}^2 e^2}{c^2 \beta^2} = 
     \frac{4 \pi n_\mathrm{e} e^4}{ m_\mathrm{e} c^2 \beta^2} = 
     \frac{3}{2} \frac{8 \pi e^4}{3 m_\mathrm{e}^2 c^4} \frac{n_\mathrm{e} m_\mathrm{e} c^2}{\beta^2}
     = \frac{3}{2} \frac{ \sigma_\mathrm{T} n_\mathrm{e} m_\mathrm{e} c^2}{\beta^2} \,.
\end{align*}
The logarithmic term can split up as
    \begin{align*}
        \ln \left(\frac{2 m_\mathrm{e} c p \beta}{\hbar \omega_{\mathrm {pl}} m_\mathrm{p} }\right) = \ln \left(\frac{2 m_\mathrm{e} c^2}{\hbar \omega_{\mathrm {pl}} }\right) + \ln \left(\frac{p }{m_\mathrm{p} c}\, \beta\right) \approx \ln \left(\frac{2 m_\mathrm{e} c^2}{\hbar \omega_{\mathrm {pl}} }\right)\,,
    \end{align*}
    where the first term lies in the range $[35.6, 42.5]$ for typical electron densities $n_\mathrm{e} \in [10^{-4},10^2]~\si{cm^{-3}}$, whereas the second term (containing all momentum-dependent quantities) lies in the range $[-10,10]$ for the relevant momenta $p/m_\mathrm{p}c \in [10^{-2}, 10^4]$. Hence, both the momentum-dependent part of the logarithm and the term $-\beta^2/2$ in brackets can be neglected, so Eq.~\eqref{eq: bCoulomb according to Gould} simplifies to
    \begin{align*}
        \left( \dv{p}{t} \right)_\mathrm{Coulomb} &\approx -\frac{\omega_{\mathrm {pl}}^2 e^2}{ c^2 }\, \ln \left(\frac{2 m_\mathrm{e} c^2}{\hbar \omega_{\mathrm {pl}} }\right) \beta^{-2}\\
        &\approx -\frac{3 \sigma_\mathrm{T} n_\mathrm{e} m_\mathrm{e} c^2}{2}\, \ln \left(\frac{2 m_\mathrm{e} c^2}{\hbar \omega_{\mathrm {pl}} }\right) ((p/m_\mathrm{p}c)^{-2}+1) \,.
    \end{align*}
    The approximation of $\beta^{-2}$ in the second step is motivated by the asymptotic behaviour, which is $\propto p^{-2}$ in the non-relativistic regime and nearly constant in the relativistic regime \citep{Pinzke.etal2013}:
    \begin{align*}
        \beta(\hat{p})^{-2} = \frac{1 + \hat{p}^2}{\hat{p}^2} \approx \begin{cases}
           \hat{p}^{-2} \quad &\text{for } \hat{p} \ll 1\,, \\
           1 \quad &\text{for } \hat{p} \gg 1\,.
        \end{cases}
    \end{align*}
    If the prefactor in our approximation for $\dot{p}_\mathrm{Coulomb}$ is multiplied with 0.9, the deviation from the true loss rate is ${\lesssim}20\%$ for low proton energies. In terms of the dimensionless momentum we find \citep[similar to][]{Winner.etal2019, Girichidis.etal2020}
    \begin{align*}
        \dot{\hat{p}}_\mathrm{Coul,p} &= -B_\mathrm{Coul,p} (1 + \hat{p}^{-2} )\,,  \\
        B_\mathrm{Coul,p} &= 0.9\, \frac{3 \sigma_\mathrm{T} n_\mathrm{e} m_\mathrm{e} c}{2 m_\mathrm{p}}\, \ln \left(\frac{2 m_\mathrm{e} c^2}{\hbar \omega_{\mathrm {pl}} }\right) \,.
    \end{align*}

For CR electrons, the Coulomb loss rate is a bit more complicated \citep{Gould1972a}:
\begin{align*}
     \left( \dv{p}{t} \right)_\mathrm{Coulomb} &= -\frac{\omega _{\mathrm {pl}}^2 e^2}{ c^2 \beta^2}\left[ \ln \left(\frac{m_{\mathrm e} c^2\beta \sqrt{\gamma -1}}{\hbar \omega _{\mathrm {pl}}}\right) \right. \\
   & \quad \left. - \frac{\ln(2)}{2} \left(1 + \frac{2 \gamma -1}{\gamma^2} \right) + \left(\frac{\gamma-1}{4 \gamma} \right)^2 + \frac{1}{2} \right] \,.
\end{align*}
Here, the relativistic gamma-factor 
\[ \gamma = (1- \beta^2)^{-1/2} = (1 + \hat{p}^2 )^{1/2} \] 
should not be confused with the adiabatic index from previous sections. After rewriting the r.h.s as a function of momentum, a similar term-by-term analysis as above shows that a good approximation is given by \citep{Winner.etal2019}:
\begin{equation*}    
\begin{aligned}
\dot{\hat{p}}_\mathrm{Coul,e}(p) &= -B_\mathrm{Coul,e} (1 + \hat{p}^{-2})\,, \\
\quad B_\mathrm{Coul,e} &= \frac{3 \sigma _\mathrm{T} n_\mathrm{e}c}{2} \ln \mathopen {} \left(\frac{m_\mathrm{e}c^2}{\hbar \omega _\mathrm{pl}}\right) \,. 
\end{aligned}
\end{equation*}
Note that the approximated Coulomb loss rates for both protons and electrons are very similar and only the prefactors differ, namely $B_\mathrm{Coul,e} \approx m_\mathrm{p}/m_\mathrm{e} B_\mathrm{Coul,p}$. From this it is also clear why the loss time scale is much shorter for electrons for a given momentum $\hat{p}$.

\begin{figure*}
    \includegraphics[width = \textwidth]{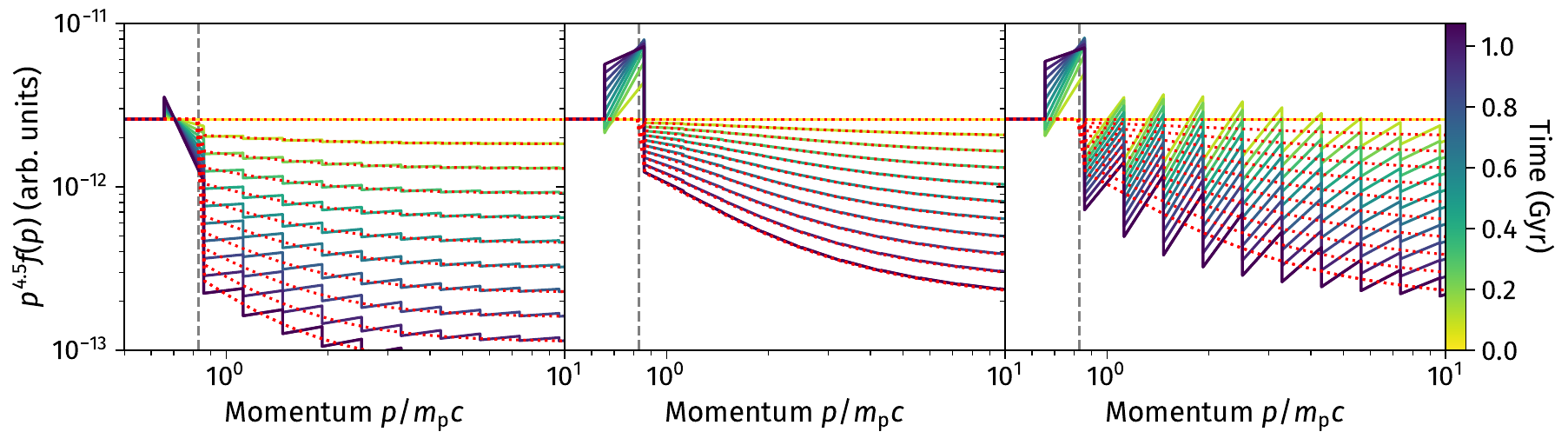}
    \caption{A CR proton spectrum is cooled via hadronic losses using three different numerical methods. This is a zoom-in close to the threshold momentum $\hat{p}_\mathrm{thr} = 0.828$ for pion production (vertical gray dashed line), where the red dotted lines are the corresponding analytical solutions from Fig.~\ref{fig: comparison between continouos and catastrophic hadronic losses}. 
    Left: Catastrophic approach~\eqref{eq: analytical solution for catastrophic hadronic losses} without accumulation of protons near $\hat{p}_\mathrm{thr}$, where the slope reconstruction at low momenta gets worse with time. 
    Middle: continuous approach~\eqref{eq: hadronic losses: continuous approach}, where the initial momentum $p_u$ is computed from the implicit equation~\eqref{eq: pu hadronic implicit}. Note that over time the protons pile up in the bin containing $\hat{p}_\mathrm{thr}$. 
    Right: continuous approach, where $p_u$ is approximated by Eq.~\eqref{eq: analytical approximation for pu_hadronic}, which is inconsistent with the loss rate~\eqref{eq: hadronic losses: continuous approach}. Hence, the slope reconstruction fails already after a short time.}
    \label{fig: simulation of 3 different approaches for hadronic losses}
\end{figure*}

\subsubsection{Bremsstrahlung losses}

In a fully ionized plasma, the momentum loss rate of a CR electron due to electron-electron and electron-nucleus bremsstrahlung was derived by \citet{Blumenthal.Gould1970}:
\begin{align}
    -\left( \dv{p}{t} \right)_\mathrm{brems} = \frac{3 \alpha_\mathrm{ fs} \sigma _\mathrm{ T} m_\mathrm{e}c^2 \bar{g}}{2 \pi}\,\frac{\gamma}{\beta}\,  \, \sum _{Z}\, n_\mathrm{ Z}\, Z\, (Z+1) \,.
\end{align}
Here, $\alpha_\mathrm{fs}$ is the fine structure constant, $n_Z$ is the number density of ions with charge number $Z$ and the sum is over all ion species (here we only consider a H-He plasma with helium mass fraction $Y$). The Gaunt factor $\bar{g} = \left(\ln(2\gamma)-1/3 \right)$ only has a weak logarithmic energy dependence, which we neglect as a lowest order approximation. Since the bremsstrahlung cooling time is shortest (and thus most relevant) for $\gamma \gtrsim 100$, we use the fixed value $\bar{g}(\gamma{=}200) \approx 5.7$, so
\begin{equation*}    
\begin{aligned}
    \dot{\hat{p}} &= - \frac{3 \alpha_\mathrm{ fs} \sigma _\mathrm{ T} c \bar{g}}{2 \pi}\, \frac{1+ \hat{p}^2}{\hat{p}}\,  (2 n_\mathrm{H} + 6 n_\mathrm{He})\\
    &= -B_\mathrm{brems} (\hat{p} + \hat{p}^{-1}) \,, \quad B_\mathrm{brems}= \frac{3 \alpha _\mathrm{ fs} \sigma _\mathrm{ T} c \bar{g}}{2 \pi} n_\mathrm{H} \frac{4 -Y}{2(1-Y)} \,, 
\end{aligned}
\end{equation*} 
where we eliminated the helium number density using
\[ n_\mathrm{He} = \frac{Y}{4(1-Y)} n_\mathrm{H}\,, \quad n_\mathrm{e} = \frac{2-Y}{2 (1-Y)} n_\mathrm{H} \,. \]
Alternatively, the coefficient $B_\mathrm{brems}$ can also be written in terms of the electron number density:
\[ B_\mathrm{brems}= \frac{3 \alpha_\mathrm{ fs} \sigma _\mathrm{ T} c \bar{g}}{2 \pi} n_\mathrm{e} \frac{4 -Y}{2-Y} \,. \]

\subsubsection{Hadronic losses}

In Sec.~\ref{sec: hadronic losses} we discussed a continuous approximation for hadronic losses, Eq.~\eqref{eq: hadronic losses: continuous approach}, where the initial momentum $p_u$ was computed numerically from the implicit equation~\eqref{eq: pu hadronic implicit}. As already mentioned, there are approximate solutions for both the non- and ultra-relativistic regime, namely Eq.~\eqref{eq: hadronic losses non-rel and ultra-rel limit}, that can be used to obtain analytical expressions for the maximum timestep and the cut-off. Now the question arises whether a simple interpolation between these limiting cases of the form
    \begin{equation}
    \begin{aligned}        
         \hat{p}_u (\hat{p} ,t) &\approx \begin{cases}
         \hat{p} & \mathrm{if}\; \hat{p} \leq \hat{p}_\mathrm{thr}\,, \\
             \zeta(\hat{p}) \hat{p}_\mathrm{n.r.} + (1- \zeta(\hat{p})) \hat{p}_\mathrm{u.r.}\, &\mathrm{if}\; \hat{p} > \hat{p}_\mathrm{thr} \,, 
         \end{cases} 
         \\ \zeta(\hat{p}) &= \frac{1}{1 + \hat{p}^4} \,,
    \end{aligned}
         \label{eq: analytical approximation for pu_hadronic}
    \end{equation}
could be used instead of the implicit equation~\eqref{eq: pu hadronic implicit} to find $\hat{p}_u$. Note that $\zeta(\hat{p})$ is just some smoothing function satisfying $\lim_{\hat{p} \ll 1}\zeta(\hat{p}) = 1$ and $\lim_{\hat{p} \gg 1}\zeta(\hat{p}) = 0$. It turns out that a simple Euler-forward-approximation for the non-relativistic regime
\[ \hat{p}_{u, \mathrm{n.r}}(\hat{p},t) = \hat{p} + B_\mathrm{had} \left(\sqrt{\hat{p}^2 +1} -1 \right) t\,, \]
is numerically more stable, because this function does not diverge in finite time. 

Consequently, we consider an initial power-law spectrum~\eqref{eq: power-law initial condition} that cools under the influence of hadronic losses. The analytical solution is given by Eq.~\ref{eq: analytical solution for f due to loss processes}, where we use the implicit equation~\eqref{eq: pu hadronic implicit} and the interpolation~\eqref{eq: analytical approximation for pu_hadronic} for $\hat{p}_u$. Both solutions agree well and only deviate close to the threshold momentum for pion production, as can be seen from Fig.~\ref{fig: comparison between continouos and catastrophic hadronic losses}. However, in this regime, Eq.~\eqref{eq: analytical approximation for pu_hadronic} is inconsistent with the loss rate \eqref{eq: hadronic losses: continuous approach}, which causes a poor slope reconstruction when $f(p,t)$ is evolved using the two-moment-approach (see right panel of Fig.~\ref{fig: simulation of 3 different approaches for hadronic losses}). This is the same numerical problem that we discussed in App.~\ref{app: Computation of the initial momentum for CR fluxes}, so the approximation~\eqref{eq: analytical approximation for pu_hadronic} should not be used.

A conceptually different approach is to treat hadronic interactions of CR protons as catastrophic loss process \citep[e.g.][]{Hopkins.etal2022a} with the general form~\eqref{eq: general form of df/dt for catastrophic losses}: 
\begin{equation}
    \begin{aligned}
        \dot{f}(p, t) &= -v(p) n_\mathrm{N} \sigma_\mathrm{pp}(p) f(p,t) = -\mathcal{R}(p) f(p,t) \,, \\
        \mathcal{R}(p) & \coloneqq n_\mathrm{N} c \sigma_\mathrm{pp} \frac{p/m_\mathrm{p}c}{ \sqrt{(p/m_\mathrm{p}c)^2 + 1}} H(p-p_\mathrm{thr}) \,.
    \end{aligned}
    \label{eq: hadronic losses: catastrophic approach}
\end{equation}
The analytical solution is directly given by
    \begin{align}
        f(p,t) = f_\mathrm{ini}(p) \exp(-\mathcal{R}(p) t)\,,
        \label{eq: analytical solution for catastrophic hadronic losses}
    \end{align}
again neglecting the weak time-dependence of $n_\mathrm{N}$. To numerically evolve the CR number and energy in each bin one does not require fluxes, but instead number- and energy-weighted loss rates:
    \begin{align*}
        \dot{n}_i &= \frac{1}{\bar{\rho}} \int \limits_{p_i}^{p_{i+1}} 4 \pi p^2 \dot{f}(p) \dd{p} 
        = - \frac{1}{\bar{\rho}} \int \limits_{p_i}^{p_{i+1}} 4 \pi p^2 \mathcal{R}(p) f(p) \dd{p} \\
        &= - \langle R \rangle_{n,i}\, n \,, \\
        \dot{\varepsilon}_i &= \frac{1}{\bar{\rho}} \int \limits_{p_i}^{p_{i+1}} 4 \pi p^2 T(p) \dot{f}(p) \dd{p} 
        = - \frac{1}{\bar{\rho}} \int \limits_{p_i}^{p_{i+1}} 4 \pi p^2 \mathcal{R}(p) T(p) f(p) \dd{p} \\
        &= - \langle R \rangle_{\varepsilon,i} \, \varepsilon \,.
    \end{align*}
    With these, the solutions are straightforward:
    \begin{align*}
        n_i(t_0 + \Delta t) &= n_i(t_0) \exp(-\langle \mathcal{R} \rangle_{n,i} \Delta t) \,, \\
        \varepsilon_i(t_0 + \Delta t) &= \varepsilon_i(t_0) \exp(-\langle \mathcal{R} \rangle_{\varepsilon,i} \Delta t) \,.
    \end{align*}
Since for the catastrophic approach protons are completely destroyed by assumption, the normalization of $f(p,t)$ decreases faster than for the continuous case, as shown in the left panel of Fig.~\ref{fig: comparison between continouos and catastrophic hadronic losses}. Consequently, if we used a smaller pion-production cross section, then it would better agree with the continuous approach. This can easily be seen in the high-momentum regime $\hat{p} \gg 1$, where the approximate loss rate $\dot{\hat{p}} = - B_\mathrm{had} \hat{p}$ and the initial condition~\eqref{eq: power-law initial condition} lead to the analytical solution
    \begin{align*}
        f_\mathrm{cont}(p,t) &= \exp(3B_\mathrm{had}t)\, f_\mathrm{ini}(p \exp(B_\mathrm{had}t)) \\
        &= f_0 p_\mathrm{min}^{q_\mathrm{ini}}\, \exp((3-q_\mathrm{ini})B_\mathrm{had}t)\, p^{-q_\mathrm{ini}}\,,
    \end{align*}
    for continuous losses, which differs from the catastrophic approach
    \begin{align*}
        f_\mathrm{cat}(p,t) = f_\mathrm{ini}(p) \exp(-2B_\mathrm{had}t) 
        = f_0 p_\mathrm{min}^{q_\mathrm{ini}}\, \exp(-2B_\mathrm{had}t) p^{-q_\mathrm{ini}} \,.
    \end{align*}
However, both solutions would coincide if we artificially rescaled the prefactor $B_\mathrm{had}$, or equivalently the cross section, to 
\begin{equation}
    \sigma_\mathrm{cat} = (q_\mathrm{ini}-3) \sigma_\mathrm{pp}/2\,.
    \label{eq: rescaled cross section for hadronic losses}
\end{equation}
As expected, this is smaller than the true cross section $\sigma_\mathrm{pp}$ for typical values of $q_\mathrm{ini}$. Nevertheless, we emphasize that for ${p} \lesssim 1$ even the rescaled solution $f_\mathrm{cat}(p,t)$ still differs noticeably from $f_\mathrm{cont}(p,t)$, which is shown in the right panel of Fig.~\ref{fig: comparison between continouos and catastrophic hadronic losses}.

Besides the different analytical solutions, there are also some numerical differences between the two approaches. To illustrate this, we plot the evolution of an initial power-law distribution function for three different implementations of hadronic losses, namely 1) the catastrophic approach~\eqref{eq: hadronic losses: catastrophic approach} and the continuous approach, where the initial momentum $p_u$ was computed using 2a) the implicit equation~\eqref{eq: pu hadronic implicit} and 2b) the explicit approximation~\eqref{eq: analytical approximation for pu_hadronic}. An important numerical effect for the latter two implementations occurs close to the threshold momentum $p_\mathrm{thr}$, which is far more likely to lie within a certain bin and not exactly at one of its boundaries. Hence, the protons with $p > p_\mathrm{thr}$ will lose their momentum and accumulate in that bin, as shown in the middle and right panel of Fig.~\ref{fig: simulation of 3 different approaches for hadronic losses}. In realistic simulations this is no big problem, because Coulomb losses at $p_\mathrm{thr}$ are typically efficient enough to remove the accumulated protons, as shown in Fig.~\ref{fig: Coulomb and hadronic losses for protons}. 

For the catastrophic approach (left panel of Fig.~\ref{fig: simulation of 3 different approaches for hadronic losses}) protons do not pile up in that bin and the steep slope is merely a numerical artifact that would not occur if $p_\mathrm{thr}$ was located at a bin boundary. However, the slope reconstruction in bins with $p \gtrsim p_\mathrm{thr}$, where $f(p,t)$ declines fastest, becomes worse with time. This is no severe problem, because 1) the reconstructed spectrum becomes significantly discontinuous only after several 100~Myrs, which is long compared to other processes that constantly reshape it in cosmological simulations, and 2) the average of the numerical spectrum is still well aligned with the analytical solution. Note that this accumulating error for the catastrophic description is fundamentally different from the issue occurring when $p_u$ is approximated by Eq.~\eqref{eq: analytical approximation for pu_hadronic}, as shown in the right panel of Fig.~\ref{fig: simulation of 3 different approaches for hadronic losses}. In this case we have an inherent inconsistency when the CR energies are updated, so the slope reconstruction according to Eq.~\eqref{eq: update of spectral slope} fails already after a short time. This is the same effect we described for Coulomb losses in Fig.~\ref{fig: Coulomb losses for CRe with pu computed in different ways}.

\section{Timesteps and number of subcycles for idealized simulations}\label{app: Comparison of timesteps for idealized simulations}
\begin{figure*}[ht]
    \centering
    \includegraphics[width=0.9\textwidth]{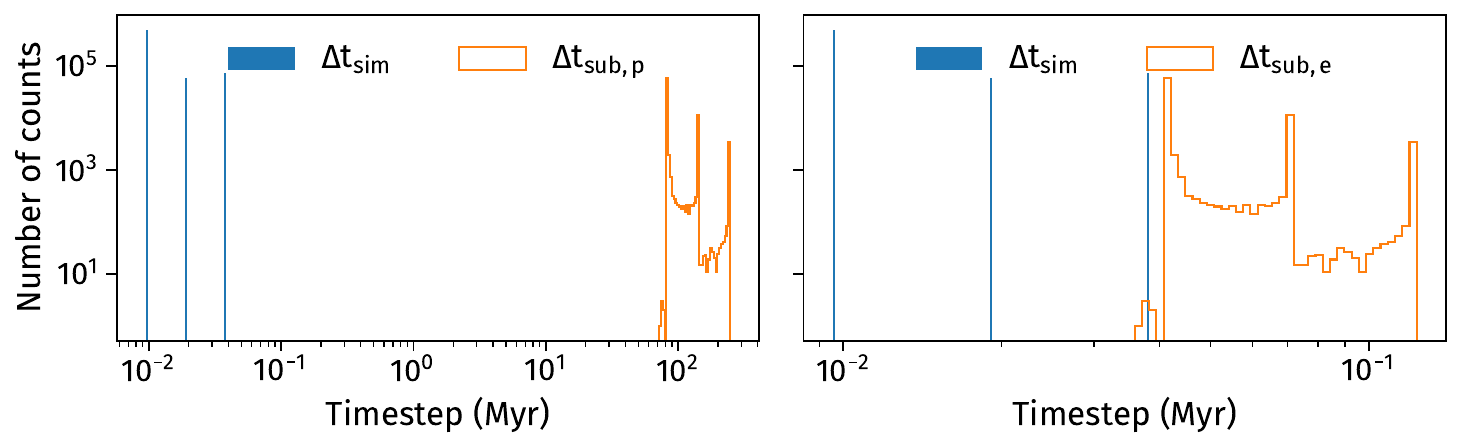}
    \caption{Simulation timesteps $\Delta t_\mathrm{sim}$ and subcycling timesteps $\Delta t_\mathrm{sub}$ for all CR protons (left) and electrons (right) in the Sod shocktube resembling typical ICM conditions. We used the same snapshot as in Fig.~\ref{fig: shocktube} showing the shock after 10~Myr and only plotted subcycling timesteps for those SPH particles with a non-zero CR population. Note that for CR protons, the values of $\Delta t_\mathrm{sub}$ are much larger than the total simulation time of 20~Myr.}
    \label{fig: timesteps for Sod shocktube}
\end{figure*}

\begin{figure*}
    \centering
    \includegraphics[width=0.9 \textwidth]{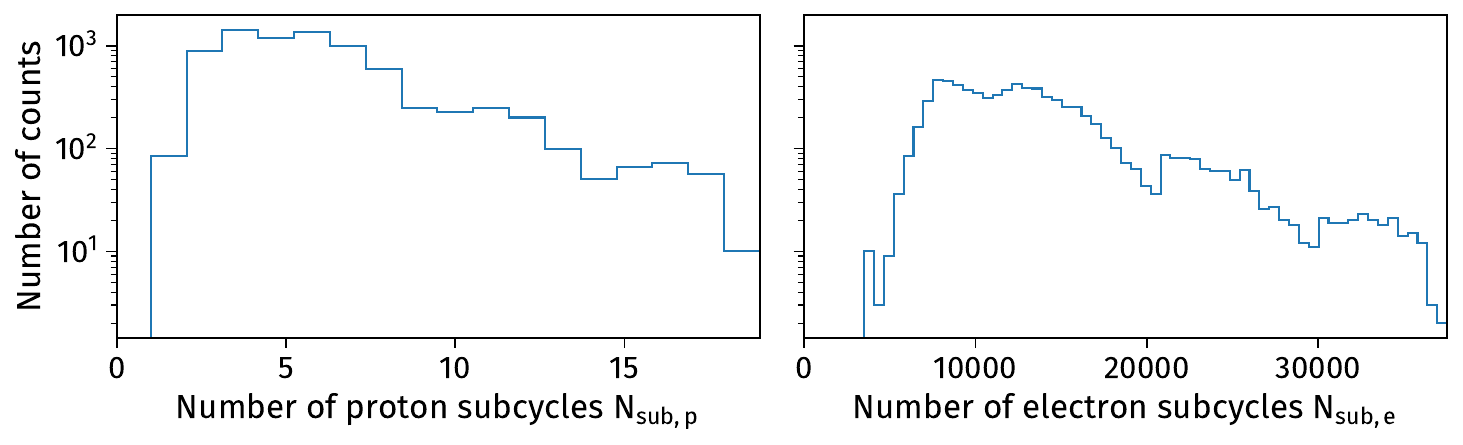}
    \caption{Histograms of the number of subcycles $N_\mathrm{sub,p}$ and $N_\mathrm{sub,e}$ for CR protons (left) and for CR electrons (right) in the ISM box described in Sec.~\ref{sec: CR seeding by supernova remnats} after 120~Myr. Note that the minimum number of electron subcycles is already $N_\mathrm{sub,e}=3501$.}
    \label{fig: subcycling histograms for ISM box}
\end{figure*}

\begin{figure*}
    \centering
    \includegraphics[height=4.7cm]{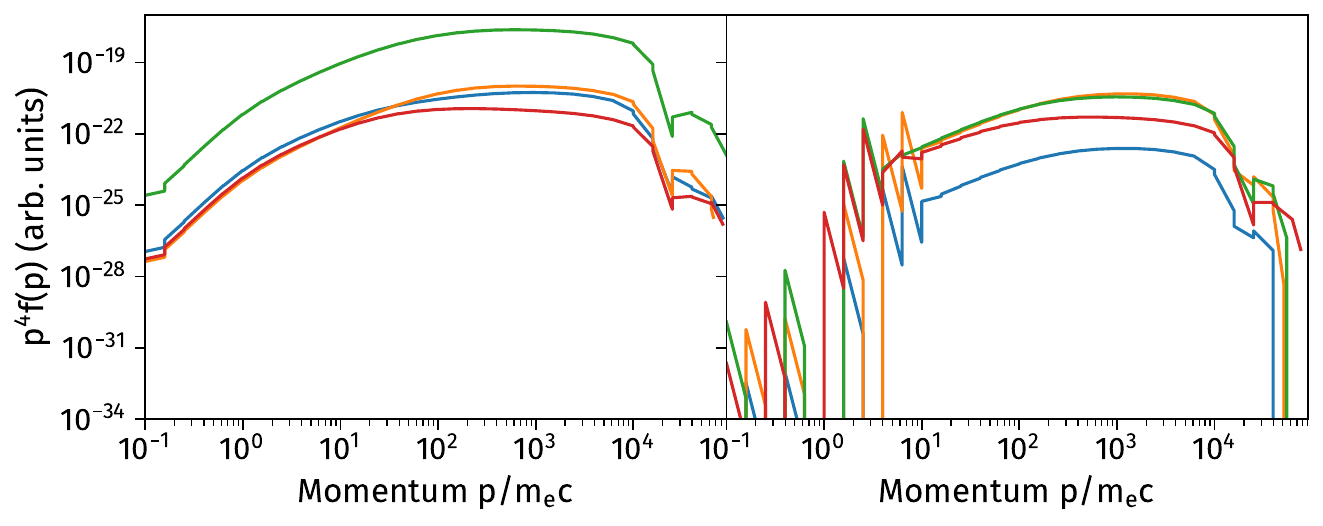}
    \includegraphics[height=4.7cm]{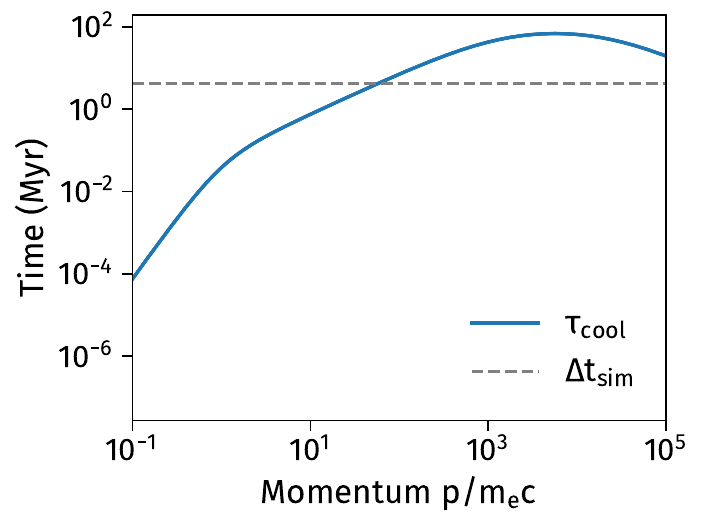}
    \caption{Electron spectra in the ISM box from Sec.~\ref{sec: CR seeding by supernova remnats} after 120~Myr. We show distribution functions of the four SPH particles with the largest number of subcycles $N_\mathrm{sub,e}$, i.e. with the shortest cooling times. In the left plot we allowed adequate subcycling of the simulation timestep, whereas in the middle plot we evolved the spectra with the much larger hydrodynamical simulation timestep. As expected, the reconstruction of the spectra is very poor in the latter case. The rightmost plot shows the cooling time $\tau_\mathrm{cool} = |p /\dot{p}|$ for the densities and magnetic fields of these four SPH particles (solid line), as well as the hydrodynamic simulation timestep $\Delta t_\mathrm{sim}$ (gray dashed line). Note that $\tau_\mathrm{cool}$ and $\Delta t_\mathrm{sim}$ are virtually identical for the selected particles.} 
    \label{fig: electron spectra and cooling times for ISM box}
\end{figure*}

As discussed in Sec.~\ref{sec: Criterion for the simulation timestep}, a CR spectrum should not be evolved over a timestep that is longer than the subcycling timestep $\Delta t_\mathrm{sub}$ from Eq.~\ref{eq: computation of Delta t_sub}, which is determined by the shortest momentum loss timescale. To explore what this means in practice, we saved both $\Delta t_\mathrm{sub}$ and the simulation timestep $\Delta t_\mathrm{sim}$ for each SPH particle when we ran the shocktube test and the ISM box from Sec.~\ref{sec: Advanced setups with multiple particles}.
In Fig.~\ref{fig: timesteps for Sod shocktube} we show the distributions of $\Delta t_\mathrm{sim}$ and $\Delta t_\mathrm{sub}$ for the shocktube test. Note that $\Delta t_\mathrm{sim}$ is always a power of 2 due to the internal timestep hierarchy of \textsc{OpenGadget3} \citep[cf.][]{2026arXiv260917658D} and $\Delta t_\mathrm{sub}$ is only plotted for particles with a non-zero CR population, where the shock wave has already passed through. It turns out that $N_\mathrm{sub} = 1$ for all particles, where the number of subcycles is defined as $N_\mathrm{sub} = \lceil \Delta t_\mathrm{sim} / \Delta t_\mathrm{sub} \rceil$. Note that the overlap between $\Delta t_\mathrm{sim}$ and $\Delta t_\mathrm{sub}$ for a small subset of CR electrons is only apparent, because the corresponding simulation timesteps are actually shorter. Moreover, for CR protons the shortest cooling time is orders of magnitude larger than the total simulation time of 20~Myr.

For the ISM box, we directly plot histograms of $N_\mathrm{sub}$ in Fig.~\ref{fig: subcycling histograms for ISM box}. Here we have the opposite situation, where $N_\mathrm{sub,p} > 10$ for a fraction of CR protons and $N_\mathrm{sub,e} > 10^4$ for the majority of CR electrons. Hence, switching off the subcycling makes the simulation ${\sim}450$ times faster, but also leads to an incorrect reconstruction of the spectrum. We show this in Fig.~\ref{fig: electron spectra and cooling times for ISM box} for those four SPH particles with the most subcycling steps. We note that this numerical experiment only works because CR numbers and energies are set to zero if the fluxes have non-finite values and the spectral slope is set to its maximum value if the slope solver does not converge. As expected, the momentum range where the cooling time $\tau_\mathrm{cool}(p)$ is shorter than the simulation timestep coincides with the range where the spectrum cannot be reconstructed anymore. Evidently, the short Coulomb timescale for CR electrons with small momenta sets the value of $N_\mathrm{sub}$ for all bins. We finally note that we used the basic expression $\tau_\mathrm{cool}(p) = \abs{p /\dot{p}}$ for the cooling time, instead of evaluating Eq.~\eqref{eq: computation of Delta t_sub} for every bin. This shows that for real cosmological simulations it is sufficient to estimate the cooling time in the densest regions based on $\tau_\mathrm{cool}$ to get an estimate for the required number of subcycles.

\section{Consistency check for the total energy of CR protons}
\label{app: The energy of CR protons is compatible with the star formation rate}

\begin{figure}
    \centering
    \includegraphics[width=\linewidth]{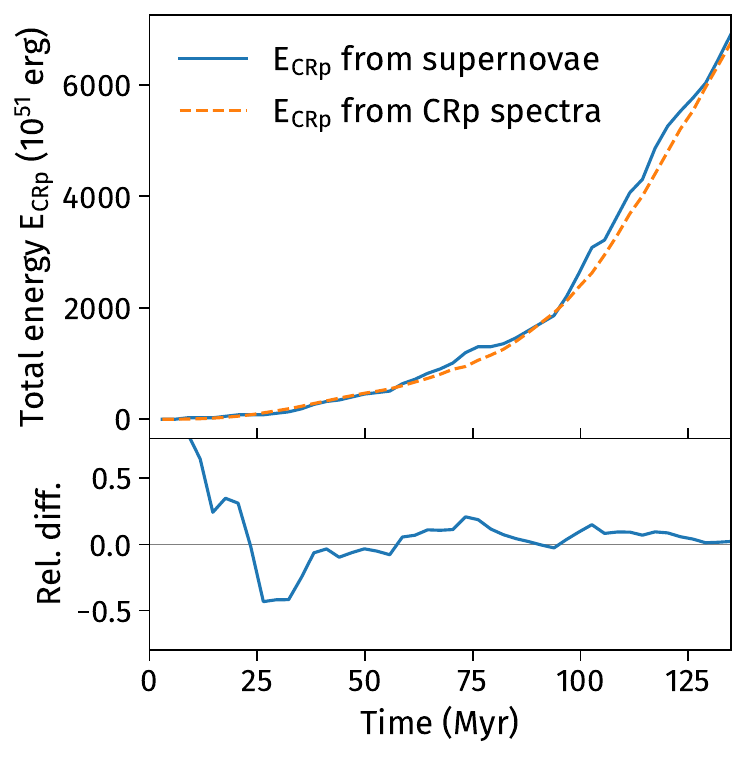}
    \caption{Top panel: Total energy of CR protons in a periodic box with ongoing star formation. The energy that is actually injected by the seeding of cosmic rays (dashed line) agrees well with what we expect from the supernova activity (solid line). Both energies increase monotonically over time, because star formation is active throughout the simulation and all loss processes for CRs are switched off. Bottom panel: Relative difference between the CR proton energies computed in the two different ways. The small fluctuations mainly arise from the stochastic nature of the effective star formation model, especially at early times when only a few stars have formed.}
    \label{fig: sfr_energies_over_time}
\end{figure}

In Sec.~\ref{sec: CR seeding by supernova remnats}, we presented an idealized setup of a periodic ISM box with ongoing star formation, where supernovae inject cosmic rays into gas particles. Here, we modify this test to check that the amount of energy that goes into CR protons is compatible with the number of supernovae occurring. The main purpose is to have a simple test for verifying that the template CR spectra are tabulated in the correct format. 

First of all, we briefly discuss how the amount of newly formed stars can be converted into feedback energy by supernovae. For this we need the initial mass function $\phi(m)$ of the stellar population, which is defined as $\phi(m) \coloneqq \dv{N_\star}{m}$, i.e. the number of stars $N_\star$ and their total mass $M_\star$ in a certain mass range $[m_1,m_2]$ are given by
\begin{align}
    N_\star = \int \limits_{m_1}^{m_2} \phi(m) \dd{m} \,, \qquad M_\star = \int \limits_{m_1}^{m_2} \phi(m) m \dd{m} \,.
\end{align}
For our test, we use the Salpeter-IMF, i.e. a power-law with a single slope, $\phi(m) \propto m^{-(1+\alpha)}$, with $\alpha=1.35$ \citep{Salpeter1955}. Moreover, all stellar masses are in the interval $[m_l,m_u] = [0.1,100]~\msol$. According to the effective star formation model of \textsc{OpenGadget3} only stars in the mass range $[m_\mathrm{SN}, m_\mathrm{BH}]$ explode as supernovae, because below $m_\mathrm{SN} = 8~\msol$ their mass is too small, whereas above $m_\mathrm{BH} = 40~\msol$ they directly collapse into a black hole by assumption. Then the mass fraction of stars undergoing supernova explosions is
\begin{align}
    \frac{M_{\star, \mathrm{SN}}}{M_{\star, \mathrm{tot}}} = \frac{\int_{m_\mathrm{SN}}^{m_\mathrm{BH}} \phi(m) m \dd{m}}{\int_{m_l}^{m_u} \phi(m) m \dd{m}} = \frac{m^{1-\alpha}_\mathrm{BH} - m^{1-\alpha}_\mathrm{SN}}{m^{1-\alpha}_u - m^{1-\alpha}_l} \approx 0.102 \,.
\end{align}
Since we assume that every supernova has an explosion energy of $E_\mathrm{SN} = 10^{51}$~erg, the released SN energy per newly formed stellar mass is given by
\begin{equation}
    \begin{aligned}
     \frac{E_{\star, \mathrm{SN}}}{M_{\star, \mathrm{tot}}} &= \frac{E_\mathrm{SN }\int_{m_\mathrm{SN}}^{m_\mathrm{BH}} \phi(m)  \dd{m}}{\int_{m_l}^{m_u} \phi(m) m \dd{m}}\\
     &= E_\mathrm{SN} \frac{\alpha -1}{\alpha} \frac{m^{-\alpha}_\mathrm{BH} - m^{-\alpha}_\mathrm{SN}}{m^{1-\alpha}_u - m^{1-\alpha}_l} \approx 6.80 \cdot 10^{48}~\mathrm{erg/\msol}\,.
\end{aligned}
\label{eq: supernova energy per stellar mass}
\end{equation}

After these introductory remarks, we now return to our idealized setup, which is the same as in Sec.~\ref{sec: CR seeding by supernova remnats}. However, this time supernovae inject simple power-law spectra $f(p) = f_0 (p/p_\mathrm{min})^{-q}$ with a single slope $q=4.3$ and normalization $f_0$ such that the integrated kinetic energy of CR protons is 10\% of the canonical SN energy $E_\mathrm{SN} = 10^{51}~\mathrm{erg}$. By choosing this ``artificial spectrum'' we know exactly which fraction of the supernova energy goes into CR protons. Hence, for our consistency check we compute the total kinetic energy of CR protons (summed over all gas particles within the box) for every simulation snapshot in two different ways. First, we know the mass of newly formed stars for each timestep and can convert this into the amount of supernova feedback energy with Eq.~\eqref{eq: supernova energy per stellar mass}. The sought-after CR proton energy should be 10\% of this. Second, we compute the energy of CR protons directly from all spectra stored in the snapshot files using Eq.~\eqref{eq: CR energy integral}. Fig~\ref{fig: sfr_energies_over_time} shows that the CR energy content of the proton spectra agrees very well with what we expect from the star formation rate.

\section{Superposition of piecewise power-law spectra} \label{app: Superposition of piecewise power-law spectra}

Suppose we have a simulation snapshot with $N_\mathrm{part}$ gas particles denoted by the index $k$ and $N_\mathrm{bins}$ momentum bins denoted by the index $i$. Then the spectrum of the $k$-th particle is represented as (cf. Eq.~\eqref{eq: piecewise power-law representation of f})
\begin{align*}
    f_{k}(p) = f_{k,i} \left( \frac{p}{p_i} \right)^{-q_{k,i}}\,, \; i \in \{1, \dots, N_\mathrm{bins}\}\,, \; k \in \{1, \dots, N_\mathrm{part}\}\,.
\end{align*}
The superposition of all spectra,
\begin{align*}
    f_\mathrm{tot}(p) = \sum_{k=1}^{N_\mathrm{part}} f_k(p)\,,
\end{align*}
can be approximated by another piecewise power-law
\begin{align*}
    f_\mathrm{tot}(p) &= f_{\mathrm{tot},i} \left( \frac{p}{p_i} \right)^{-q_{\mathrm{tot},i}}\,, \; i \in \{1, \dots, N_\mathrm{bins}\}\,.
\end{align*}
If one explicitly evaluates $f_\mathrm{tot}(p=p_i)$ and $f_\mathrm{tot}(p=p_{i+1})$, then the total normalizations $f_{\mathrm{tot},i}$ and slopes $q_{\mathrm{tot},i}$ can be written in terms of the known quantities $f_{k,i}$ and $q_{k,i}$:
\begin{align*}
    f_{\mathrm{tot},i} &= \sum_{k=1}^{N_\mathrm{part}} f_{k,i} \,, \\
    q_{\mathrm{tot},i} &= - \lg\left( \frac{\sum_{k=1}^{N_\mathrm{part}} f_{k,i} \left(p_{i+1}/p_i \right)^{-{q_{k,i}}}}{\sum_{k=1}^{N_\mathrm{part}} f_{k,i}} \right) \bigg/ \lg\left( \frac{p_{i+1}}{p_i} \right) \,.
\end{align*}
This superposition is shown in Figs.~\ref{fig: injection of CR spectra} and \ref{fig: CR spectrum at the shock}, where the logarithmic derivative of the CR kinetic energy $E_\mathrm{CR}$ is given by
\begin{align*}
    \dv{E_\mathrm{CR}}{\lg(\hat{p})} &= \frac{p}{\lg(e)}\, \dv{E_\mathrm{CR}}{p} =  \frac{p}{\lg(e)}\, 4\pi p^2 T(p) f(p)\\
    &= \frac{4 \pi (mc)^4c}{\lg(e)} \hat{p}^3 \left(\sqrt{\hat{p}^2 + 1} - 1 \right) f(\hat{p})\,.
\end{align*}

\FloatBarrier 
\clearpage

\end{appendix}
\end{document}